\documentclass{aa}  
\usepackage{graphicx}
\usepackage{txfonts}
\usepackage{natbib}
\bibpunct{(}{)}{;}{a}{}{,} 
\newcommand{\mincir}{\raise -2.truept\hbox{\rlap{\hbox{$\sim$}}\raise5.truept
\hbox{$<$}\ }}
\newcommand{\magcir}{\raise -2.truept\hbox{\rlap{\hbox{$\sim$}}\raise5.truept
\hbox{$>$}\ }}
\newcommand{\siml}{\raise -2.truept\hbox{\rlap{\hbox{$\sim$}}\raise5.truept
\hbox{$<$}\ }}
\newcommand{\simg}{\raise -2.truept\hbox{\rlap{\hbox{$\sim$}}\raise5.truept
\hbox{$>$}\ }}
\newcommand{\be}{\begin{equation}}
\newcommand{\ee}{\end{equation}}
\newcommand{\ba}{\begin{eqnarray}}
\newcommand{\ea}{\end{eqnarray}}
\newcommand {\kpc} {kpc $\;$}

\newcommand {\h} {Mpc$\;$}
\newcommand {\hh} {Mpc}

\newcommand {\ks} {km~s$^{-1} \;$}
\newcommand {\kss} {km~s$^{-1}$}

\newcommand {\mqua} {$\times 10^{14}\;M_{\odot} \;$}
\newcommand {\mquaa} {$\times 10^{14}\;M_{\odot}$}
\newcommand {\mqui} {$\times 10^{15}\;M_{\odot} \;$}
\newcommand {\mquii} {$\times 10^{15}\;M_{\odot}$}
\newcommand {\ml} {$M_{\odot}/L_{\odot} \;$}
\newcommand {\mll} {$M_{\odot}/L_{\odot}$}

\newcommand{\degree}{\ensuremath{\mathrm{^\circ}}}
\newcommand{\arcm}{\ensuremath{\mathrm{^\prime}\;}}
\newcommand{\arcs}{\ensuremath{\arcmm\hskip -0.1em\arcmm \;}}
\newcommand{\arcmm}{\ensuremath{\mathrm{^\prime}}}

\newcommand{\dotsec}{\,\rlap{\hbox{$\mathrm{^s}$}}{\hbox{$.$}}\,}

\begin{document}
\title{The velocity field of the Lyra complex}  

%

   \author{
M. Girardi\inst{1,2},
          W. Boschin\inst{3,4,5},
          S. De Grandi\inst{6}, M. Longhetti\inst{6}, S. Clavico\inst{6,7}, D. Eckert\inst{8}, F. Gastaldello\inst{9}, S. Ghizzardi\inst{9},  M. Nonino\inst{2}, M. Rossetti\inst{9}}


  \institute{Dipartimento di Fisica dell'Universit\`a degli Studi di Trieste -
Sezione di Astronomia, via Tiepolo 11, I-34143 Trieste, Italy \email{marisa.girardi@inaf.it}
\and INAF - Osservatorio Astronomico di Trieste, via Tiepolo 11,
I-34143 Trieste, Italy
\and Fundaci\'on Galileo Galilei - INAF (Telescopio Nazionale
  Galileo), Rambla Jos\'e Ana Fern\'andez Perez 7, E-38712 Bre\~na
  Baja (La Palma), Canary Islands, Spain
\and Instituto de Astrof\'{\i}sica de Canarias, C/V\'{\i}a L\'actea
s/n, E-38205 La Laguna (Tenerife), Canary Islands, Spain
\and Departamento de Astrof\'{\i}sica, Univ. de La Laguna, Av. del
Astrof\'{\i}sico Francisco S\'anchez s/n, E-38205 La Laguna
(Tenerife), Spain
\and INAF - Osservatorio Astronomico di Brera, via E. Bianchi 46, I-23807 Merate, Italy
\and Dipartimento di Fisica dell'Universit\`a Bicocca di Milano -
Milano, Italy
\and Department of Astronomy, University of Geneva, ch. d'Ecogia 16, 1290 Versoix, Switzerland
\and INAF - IASF Milano, via E. Bassini 15, 20133, Milano, Italy
  }

\date{Received  / Accepted }

\abstract {The formation of cosmic structure culminates with
  the assembly of galaxy clusters, a process quite different from
  cluster to cluster.}  {We present the study of the structure and
  dynamics of the Lyra complex formed of the two clusters
  RXC~J1825.3+3026 and CIZA~J1824.1+3029, very recently studied using
  both X-ray and radio data.}  {This is the first analysis based on
  kinematics of member galaxies. New spectroscopic data for 285
  galaxies were acquired at the Italian Telescopio Nazionale {\em
    Galileo} and used in combination with PanSTARRS photometry. The
  result of our member selection is a sample of 198 galaxies.}  {For
  RXCJ1825 and CIZAJ1824 we report the redshifts, $z=0.0645$ and
  $z=0.0708$, the first estimates of velocity dispersion, $\sigma_{\rm
    v}=995_{-125}^{+131}$ \ks and $\sigma_{\rm v}=700\pm50$ \kss, and
  dynamical mass, $M_{\rm 200}=1.1\pm 0.4$ \mqui and $M_{\rm 200}=4\pm
  0.1$ \mquaa.  The past assembly of RXCJ1825 is 
  traced by the two dominant galaxies, both aligned
  with the major axis of the galaxy distribution
  along the East--West direction,  and by a minor North--East
  substructure. We also detect a quite peculiar high velocity field in
  the South--West region of the Lyra complex. This feature is likely
  related to a high velocity, very luminous galaxy, suggested to be
  the central galaxy of a group in interaction with RXCJ1825
  by very recent studies based on X-ray and radio data.  The
  redshift of the whole Lyra complex is $z=0.067$.  Assuming that the
  redshift difference between RXCJ1825 and CIZAJ1824 is due to the
  relative kinematics, the projected distance between the cluster
  centers is $D\sim 1.3$ \h and the line--of--sight velocity
  difference is $\sim 1750$ \kss.  A dynamical analysis of the system
  shows that the two clusters are likely to be gravitationally bound, in a
  pre-merger phase, with CIZAJ1824 in front of RXCJ1825 and going
  toward it.}{Our results corroborate a picture where the Lyra region
  is the place of a very complex scenario of cluster assembly.}

\titlerunning{Clusters RXCJ1825 and CIZAJ1824} 
\authorrunning{Girardi et al.} 
   \maketitle

\section{Introduction}
\label{intr}

Clusters of galaxies are the largest gravitationally bound systems in
the Universe.  According to the $\Lambda$CDM hierarchical scenario,
the formation of structure progresses in a hierarchical fashion,
culminating with the assembly of clusters of galaxies (see
\citealt{springel2006} and refs. therein). Numerical simulations also
show that clusters form preferentially through anisotropic accretion
along the large-scale structure filaments (e.g.,
\citealt{colberg1999}) and in a significant part through the accretion
of galaxy groups, while the merger of two or more several massive
entities is a more rare case (e.g., \citealt{berrier2009};
\citealt{mcgee2009}).  Since the cluster assembly histories show
significant variation from cluster and cluster (e.g.,
\citealt{berrier2009}), most observational studies are focused on
individual systems.

From the observational side, the trace of the cluster assembly has
been studied since a long time, through the analysis of substructure
as based on cluster galaxies (\citealt{baier1977};
\citealt{geller1982}), X-ray emitting intracluster medium -- ICM
(\citealt{jones1999}) and more recently on gravitational lensing
effects (e.g, \citealt{Athreya2002}; \citealt{Dahle2002}).  Merging
clusters have proved to be fruitful laboratories to study several
physical processes.  Cluster mergers have been suggested to be the
energetic support for radio halos and relics (\citealt{tribble1993};
\citealt{feretti1999}).  The analysis of the merging system named
``Bullet-cluster'' has allowed to obtain a strong prove for the
existence of the dark matter -- DM -- showing the decoupling of
baryonic and DM (\citealt{markevitch2002}; \citealt{markevitch2004}).
The study of the accretion of groups onto clusters is quite timely in
the context of galaxy evolution (e.g., \citealt{olave-rojas2018} and
refs. therein) since some pre-processing of galaxies in the group
environment is expected prior to cluster formation (\citealt{zabludoff1998}),

Optical data are a powerful way to investigate the presence of
structures and the dynamics of cluster mergers (\citealt{girardi2002}
for a review).  Moreover, the photometric and spectroscopic
information about cluster galaxies is complementary to the X--ray
information since it is well known that galaxies and the ICM react on
different timescales during a merger as shown by numerical simulations
(e.g., \citealt{roettiger1997}; \citealt{springel2007}). In
particular, multi-object spectroscopy observations have allowed the
building of large samples of galaxies with measured redshifts for
individual clusters and have proven a powerful tool in understanding
cluster formation, as shown by several dedicated studies (e.g.,
\citealt{girardi2011}, DARC project; \citealt{maurogordato2011}, MUSIC
project; \citealt{owers2013} using GAMA survey;
\citealt{balestra2016}, CLASH-VLT project; \citealt{golovich2017},
$MC^2$ collaboration).

An ideal candidate where to study in details how structures grow is
the complex of the two clusters \object{RXC~J1825.3+3026} (RXCJ1825
hereafter) and \object{CIZA~J1824.1+3029} (hereafter CIZAJ1824),
hereafter named the Lyra complex.  RXCJ1825, also named
CIZA~J1825.3+3026, was discovered by ROSAT in the X-rays at galactic
latitude b=18.547 deg at z=0.0645 (\citealt{ebeling2002}). This
cluster was found to be one of the strongest and spatially resolved
source of Sunyaev-Zeldovich signal in the Planck all-sky cluster
survey (\citealt{planckVIII}). It has been studied as part of the {\em
  XMM-Newton} Cluster Outskirts Project (X-COP, \citealt{eckert2017}),
a very large program based on {\em XMM-Newton} X-ray observations
(\citealt{eckert2019}; \citealt{ettori2019}; \citealt{ghirardini2019}).  In
particular, as shown by Fig.~2 of \citet{ghirardini2019}, about
16\arcm West, slightly WNW, of this cluster there is the smaller 
cluster CIZAJ1824, already named NPM1G+30.0 and listed by
\citet{voges1999}, with redshift $z=0.072$ (\citealt{kocevski2007}).
The recent study of Clavico et al. (\citeyear{clavico2019}) shows that whereas CIZAJ1824
is dynamically relaxed, RXCJ1825 is not and shows clear signatures of
past and on-going merging. Indeed, the presence of a giant radio halo
in RXCJ1825 is likely related to the on-going merging in this system
(\citealt{botteon2019}).

For these two clusters there were no ad hoc optical observations
available.  No dynamical analysis has never been performed and indeed
only one redshift per cluster was known.  We obtained time to perform
a spectroscopic survey at the Italian Telescopio Nazionale {\em Galileo}
(TNG). This study is devoted to the presentation of our analysis of
the velocity field of the Lyra complex. The paper is organized as
follows. We present the optical data and the cluster catalogue in
Sect.~\ref{data}. In Sect.~\ref{memb} we describe the member selection
procedure and global properties of the cluster
complex. Section~\ref{sub} is devoted to the analysis of the structure
of the Lyra complex.  We report our estimates of properties of the two
individual clusters in Sects.~\ref{dise} and ~\ref{mass}.  In
Sect.~\ref{bim} we present our dynamical analysis for the whole
complex.  We discuss our results in Sect.~\ref{discu} and present our
conclusions in Sect.~\ref{summa}.

Unless otherwise stated, we indicate errors at the 68\% confidence
level (hereafter c.l.).  Throughout this paper, we use $H_0=70$ km
s$^{-1}$ Mpc$^{-1}$ in a flat cosmology with $\Omega_0=0.3$ and
$\Omega_{\Lambda}=0.7$. In the adopted cosmology, 1\arcm corresponds
to $\sim 77.5$ \kpc at the redshift of the Lyra complex, used to fix
the \h scale of projected distances throughout the paper.
All magnitudes are presented in the AB system.

\section{Data and galaxy catalog}
\label{data}

\subsection{New Spectroscopic observations}
\label{spec}

We observed the Lyra field with DOLORES@TNG in June 2018. In particular, we
made use of the multi-object spectroscopy (MOS) capabilities of this
instrument to observe 12 MOS masks with the LR-B
Grism\footnote{http://www.tng.iac.es/instruments/lrs}. In total, we
observed 12 MOS masks for a total of 390 slits. For 9 masks, the total
exposure time was 3600 s, for 3 masks the exposure time was 5400 s.
In February 2019 we took long-slit spectra of another two galaxies
(one of them is the bright radio galaxy ID.~039, see Sect.~\ref{cat}),
both observed with an exposure time of 1800s.

We performed the reduction of the optical spectra with standard
IRAF\footnote{IRAF is distributed by the National Optical Astronomy
  Observatories, which are operated by the Association of Universities
  for Research in Astronomy, Inc., under cooperative agreement with
  the National Science Foundation.} tasks and were able to compute
redshifts for 256 galaxies by using the cross-correlation method
introduced by \citet{tonry1979}. For another 29 galaxies we estimated
the redshifts by computing the wavelength location of emission lines
in their spectra (more details of the data reduction in, e.g.,
\citealt{boschin2013}).

\begin{figure*}[!ht]
\centering 
\includegraphics[width=17cm]{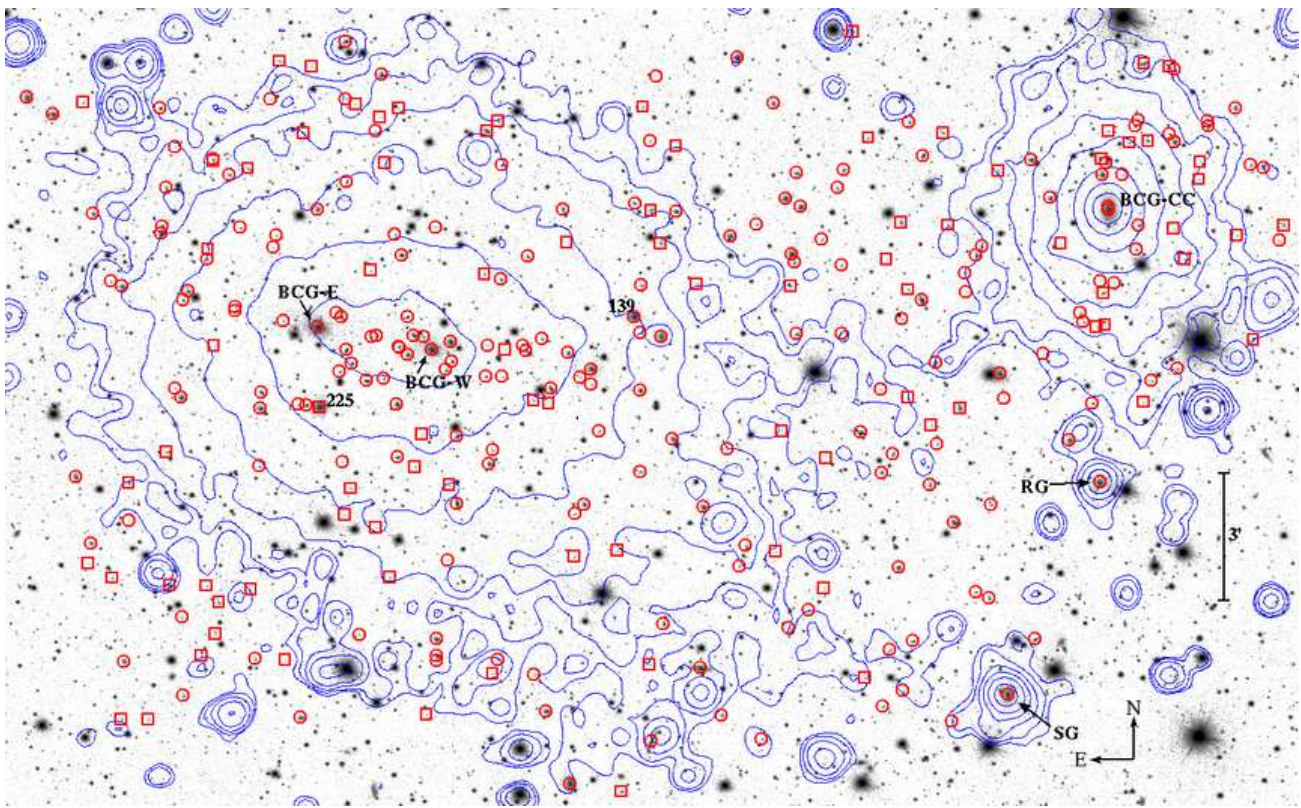}
\caption{Pan-STARRs r-band image of the Lyra complex
  (RXCJ1825+CIZAJ1824 clusters) with, superimposed, the contour levels
  of the XMM X-ray emission taken from Clavico et
  al. (\citeyear{clavico2019}).  Circles and squares indicate cluster
  members and non-members, respectively (see
  Table~\ref{catalogRXCJ1825}). Labels indicate galaxies cited in the
  text.}
\label{figottico}
\end{figure*}

In total, our spectroscopic catalog lists 285 galaxies in the field of
RXCJ1825. The median value of the $cz$ errors is 114 \kss.
Figure~\ref{figottico} shows the field of the Lyra complex sampled by
TNG spectroscopic data.

We also used the magnitudes $g$, $r$, and $i$ available in Pan-STARRS
(DR1; Chambers et al. 2016) after correction for Galactic
absorption.With the exception of 16 galaxies, $g$ magnitudes are
available. Out of these, four galaxies have not $r$ and $i$ magnitudes
available.

\subsection{Galaxy catalog and prominent galaxies}
\label{cat}

\begin{table}
        \caption[]{Velocity catalog of 285 spectroscopically measured
          galaxies in the field of the Lyra complex.}
         \label{catalogRXCJ1825}
              $$ 
           \begin{array}{r c c c r r}
            \hline
            \noalign{\smallskip}
            \hline
            \noalign{\smallskip}

\mathrm{ID} & \mathrm{Member} & \mathrm{\alpha},\mathrm{\delta}\,(\mathrm{J}2000)  & r & \mathrm{v}\,\,\,\,\,&\mathrm{\Delta}\mathrm{v}\\
  & &(18^{h},+30^{o}) &(\mathrm{mag}) &\mathrm{(\,km}&\mathrm{s^{-1}\,)}\\
            \hline
            \noalign{\smallskip}  

001 & \mathrm{N}           & 23\ 48.10 ,29\ 10.3&      18.66&  77576&  96 \\      
002 & \mathrm{Y}           & 23\ 48.47 ,28\ 48.6&      19.85&  19910& 192 \\      
003 & \mathrm{Y}           & 23\ 50.28 ,30\ 30.0&      18.03&  21914&  78 \\      
004 & \mathrm{N}           & 23\ 51.39 ,26\ 30.3&      18.66&  36005& 114 \\      
005 & \mathrm{Y}           & 23\ 51.61 ,30\ 34.2&      18.87&  21980& 177 \\      
                        \noalign{\smallskip}			    
            \hline					    
            \noalign{\smallskip}			    
            \hline					    
         \end{array}
           $$
                    \tablefoot{Full table is available at CDS}
\end{table}

Table~\ref{catalogRXCJ1825}, available at the CDS, lists the velocity
catalog (see also Fig.~\ref{figottico}): identification number of each
galaxy and member galaxies, ID and membership (Cols.~1 and 2,
respectively); right ascension and declination, $\alpha$ and $\delta$
(J2000, Col.~3); dereddened $r$ Pan-STARRS (DR1) magnitudes, $r$
(Col.~4); heliocentric radial
\footnote{Unless otherwise stated, velocities reported in the paper
  are radial, that is line--of--sight ones.}  velocities, ${\rm
  v}=cz_{\sun}$ (Col.~5) with errors, $\Delta {\rm v}$ (Col.~6).
An excerpt from this table is also inserted in the article. 

\begin{figure}
\centering
\resizebox{\hsize}{!}{\includegraphics{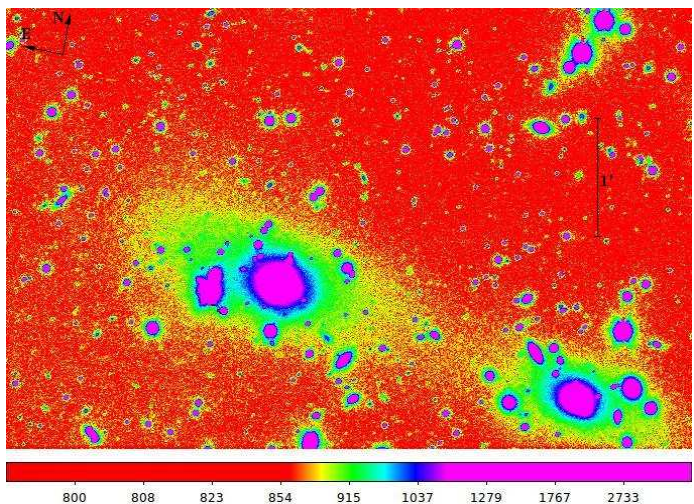}}
\caption
    {TNG V-band image of the RXCJ1825 region around the two brightest
      galaxies (BCG-E and BCG-W). The color scale units are ADU using a  logarithmic display function.
      }
\label{figBCGs}
\end{figure}

RXCJ1825 hosts two dominant galaxies, the galaxy ID.~226 ($r$=14.63,
hereafter BCG-E) and the galaxy ID.~186 ($r$=15.15, hereafter
BCG-W). The centroid of the XMM X-ray emission lies between these two
galaxies, much closer to the BCG-W. Hereafter, for the center of
RXCJ1825 we adopt the position of the X-ray centroid reported by
Clavico et al. (\citeyear{clavico2019}),
[R.A.=$18^{\mathrm{h}}25^{\mathrm{m}}21\dotsec77$, Dec.=$+30\degree
  26\arcmm 25.3\arcs$ (J2000.0)].  The two BCGs are shown in
Fig.~\ref{figBCGs}.

The companion cluster, CIZAJ1824, hosts a cool core (\citealt{clavico2019}) surrounding the dominant galaxy, the galaxy ID.~030
($r$=14.68, hereafter BCG-CC).
As for the center of CIZAJ1824,
we adopt the position of its BCG
[R.A.=$18^{\mathrm{h}}24^{\mathrm{m}}06\dotsec98$, Dec.=$+30\degree
  29\arcmm 30.4\arcs$ (J2000.0)].

At the South--West of the region, the galaxy ID.~050 ($r=15.28$,
hereafter SG) is at the top of an extended X-ray emission detected by
Clavico et al. (\citeyear{clavico2019}), see also
Fig.~\ref{figottico}.  They suggest that SG once was the central
galaxy of a group, now in an advanced state of disruption after the
interaction with RXCJ1825. This scenario explains the excess in the
X-ray surface brightness between RXCJ1825 and SG (see
\citealt{clavico2019} for details). The position of SG is
R.A.=$18^{\mathrm{h}}24^{\mathrm{m}}17\dotsec68$, Dec.=$+30\degree
18\arcmm 15.0\arcs$ (J2000.0).

Other interesting galaxies are those with a peculiar radio emission
(\citealt{botteon2019}).  The ID.~139 ($r=15.42$) is located West of
RXCJ1825 center, in direction of CIZAJ1824, showing the features of a
tailed radio galaxy. In the West region of the field, South of
CIZAJ1824, there is another much brighter tailed radio galaxy, which
coincides with the optical galaxy ID.~039 (hereafter RG).

\section{Member selection and global properties}
\label{memb}

\begin{figure}
\centering
\resizebox{\hsize}{!}{\includegraphics{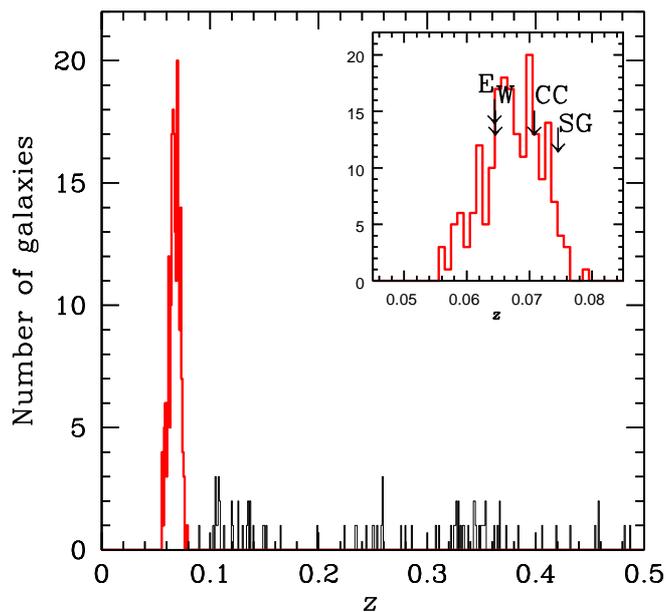}}
\caption
{Redshift galaxy distribution. The solid red line histogram refers to
  the galaxies assigned to the Lyra complex according to
  the DEDICA reconstruction method. The distribution of the 198 member
  galaxies with the redshift of prominent galaxies is
  shown in the inset plot.}
\label{fighisto}
\end{figure}

To select cluster members among the 285 galaxies with redshifts we
applied the 1D adaptive-kernel method (hereafter DEDICA,
\citealt{pisani1993}), very efficient for both medium and high sampled
fields (\citealt{fadda1996}; \citealt{balestra2016}). This method
searches for the most significant peaks in the velocity
distribution. We detected the Lyra complex as
a peak at $z\sim0.067$ populated by 199 galaxies (in the range
$0.055518\leq z \leq 0.079265$, see Fig.~\ref{fighisto}).
The non-selected galaxies are all background galaxies.

We also reject the very bright galaxy ID.~225 ($r$=15.01) which lies
close to the lower limit of our redshift selection. The inspection of
our TNG image reveals that it is a huge spiral. In the case that
  ID.~225 were part of the cluster, this bright galaxy would be the
  second brightest galaxy in the RXCJ1825 core, even more luminous
  than BCG-W.  However, in the literature non elliptical BCGs are very
  rare and in a few of these cases richer data sets have shown that
  they are cases of misclassification, that is galaxies belonging to a
  small foreground group in front of the cluster (e.g.,
  \citealt{lauer2014}). Indeed, the redshift constraints imposed by
  the member selection cannot prevent from including those galaxies
  with a redshift similar to that of the cluster but really distant
  (several Mpc) from the cluster in the three-dimensional space, as
  also shown when projecting simulated clusters and field around
  (e.g., \citealt{biviano2006}). Therefore we assume that the correct
interpretation is that ID.~225 is a foreground object at z=0.056075,
$\sim 50$ Mpc in front of the cluster complex, and projected onto the
core of RXCJ1825.

The final catalog of member galaxies
is made of 198 objects, 194 having full photometric information, too.

By applying the biweight estimator to the 198 members of the complex,
we computed a mean redshift of $\left<z\right>=0.0674\pm$ 0.0003, i.e.
$\left<\rm{v}\right>=20\,203\pm$96 \ks (\citealt{beers1990},
ROSTAT software). We estimated the velocity dispersion of the
whole complex, $\sigma_{\rm v}$, by using the biweight estimator and
applying the cosmological correction and the standard correction for
velocity errors (\citealt{danese1980}).  We obtained $\sigma_{\rm
  v}=1342_{-68}^{+60}$ \kss, where errors are estimated through a
bootstrap technique.

\section{Structure of the Lyra field}
\label{sub}

\subsection{2D structure}
\label{2D}

\begin{figure}
\resizebox{\hsize}{!}{\includegraphics{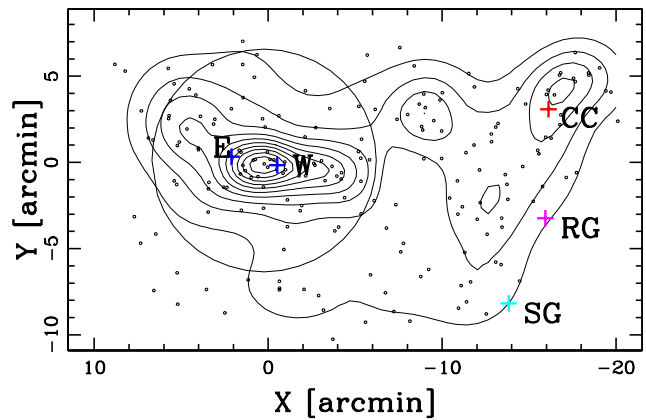}}
\caption
{Spatial distribution on the sky and relative isodensity contour map
  of the 198 spectroscopic members of the Lyra complex, obtained with the
  2D-DEDICA method. The peaks from East to West are RXCJ1825NE,
  RXCJ1825main, MiddlePeak, CIZAJ1824 (see Tab.~\ref{tabdedica2d}).
  The X-ray centroid in RXCJ1825 is taken as the center of the whole
  complex.  The position of the prominent galaxies
  (BCGs of RXCJ1825, BCG of CIZAJ1824, SG, RG) are indicated by
  crosses (blue, red, cyan, and magenta, respectively).  The region of
  RXCJ1825 within 0.5 \h is enclosed by the circle for an easier
  comprehension of the projected \h size.  }
\label{figk2z}
\end{figure}

In order to determine the structure of the Lyra complex as projected
on the plane of the sky, we applied the 2D adaptive-kernel method
(2D-DEDICA, \citealt{pisani1996}) to the positions of galaxy members,
We detected four peaks with a significance larger than the 99\%
c.l. and a relative density with respect to the densest peak $\rho_S
>$0.3 (see Fig.~\ref{figk2z} and Tab.~\ref{tabdedica2d}).  The most
significant and dense peak (RXCJ1825main) individuates the RXCJ1825
cluster, close to the X-ray centroid, and the isodensity curves of the
galaxy distribution design a structure elongated in the East--West
direction, close to that traced by the position of the two
BCGs. North--East of the cluster peak, a minor peak is also detected
(RXCJ1825NE). The second significant and dense peak individuates the
CIZAJ1824 cluster. The last peak lies between the two clusters
(MiddlePeak).

\begin{table}
        \caption[]{2D substructure from the analysis of
          spectroscopic members.}
         \label{tabdedica2d}
            $$
         \begin{array}{l r c c c }
            \hline
            \noalign{\smallskip}
            \hline
            \noalign{\smallskip}
\mathrm{Subclump} & N_{\rm S} & \alpha({\rm J}2000),\,\delta({\rm J}2000)&\rho_{
\rm S}&\chi^2_{\rm S}\\
& & \mathrm{h:m:s,\degree:\arcmm:\arcs}&&\\
         \hline
         \noalign{\smallskip}
\mathrm{RXCJ1825main} & 69&18\ 25\ 22.9+30\ 26\ 17&1.00&15\\
\mathrm{CIZA1824}     & 27&18\ 24\ 05.1+30\ 30\ 18&0.39&11\\
\mathrm{MiddlePeak}   & 20&18\ 24\ 39.9+30\ 29\ 13&0.33& 5\\
\mathrm{RXCJ1825NE}   & 28&18\ 25\ 42.7+30\ 28\ 16&0.32& 6\\
              \noalign{\smallskip}
              \noalign{\smallskip}
            \hline
            \noalign{\smallskip}
            \hline
         \end{array}
$$
         \end{table}

\begin{figure}
\centering
\resizebox{\hsize}{!}{\includegraphics{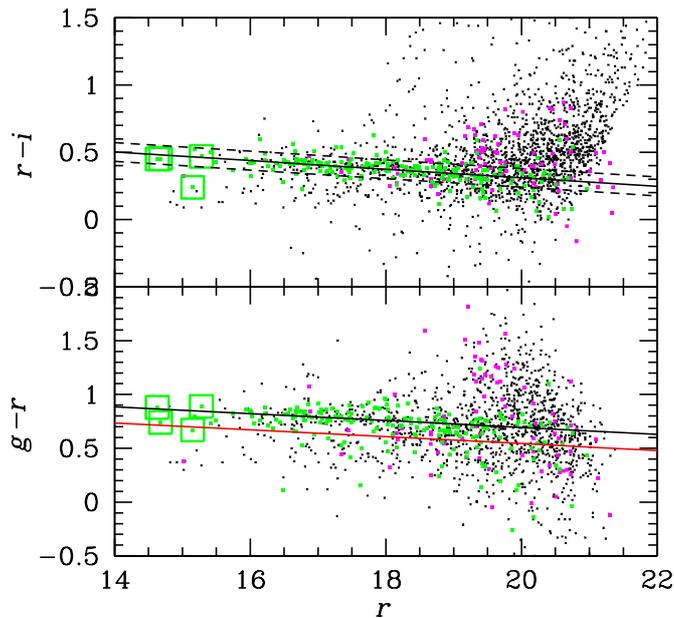}}
\caption
{{\em Upper panel:} Pan-STARRS $r-i$ vs. $r$ diagram.  Small black
  symbols indicate the galaxies of the photometric catalog, green and
  magenta symbols indicate member and no-member galaxies of our
  spectroscopic catalog, large green squares indicate prominent
  galaxies (BCG-E, BCG-CC, BCG-W, and SG in order of increasing
  $r$ magnitude).  The solid line gives the CMR determined on
  spectroscopic member galaxies; the dashed lines are drawn at
  $\pm$0.07 mag from this value and are used to define the photometric
  members in the Pan-STARRS catalog.  {\em Lower panel:} As above, but
  for the Pan-STARRS $g-r$ vs. $r$ diagram.  The red solid is drawn at
  0.15 mag down of the CMR, determined on spectroscopic member galaxies,
  and is used to discriminate red from blue galaxies.}
\label{figcm}
\end{figure}

Our spectroscopic data do not cover the entire cluster field in an
uniform way and, in particular, the position of masks might bias the
result. To check our results we used a photometric catalog extracted
from the Pan-STARRS survey.  We considered non-stellar objects within
a radius of $\sim$ 20\arcm from the central point of our spectroscopic
observations and we applied the magnitude corrections for the Galactic
absorption.  We selected likely members on the basis of the $r$--$i$
vs. $r$ color-magnitude relation (CMR), which indicate the locus of
member galaxies (e.g., \citealt{goto2002}; see Fig.~\ref{figcm}).  To
determine the CMR we applied the 2$\sigma$-clipping fitting procedure
to the cluster members: we obtained $r$--$i$=0.950-0.032$\times r$,
based on 132 surviving galaxies.  Our fitted relation agrees with
  the fact that $r$--$i$ vs. $r$ CMR is about flat
  (\citealt{goto2002}), with slope values depending on the analyzed
  cluster (e.g., between -0.04 and 0. in the sample of
  \citealt{barrena2012}).  As for the intercept, our value of
  $r$--$i$=0.39 for $r$=17.5 is well consistent with the value
  reported for low redshift clusters, $r$--$i$$\sim 0.4$ for Abell 168
  at $z\sim 0.044$ and for Abell 1577 at $z\sim$ 0.14
  (\citealt{goto2002}, see Figs.~1 and 2 and text).  Out of the
Pan-STARRS photometric catalog we consider as likely cluster members
the objects lying within 0.07 mag from the CMR, that is about two
times the error on the intercept.

  \begin{table}
        \caption[]{2D substructure from the analysis of photometric members.}
         \label{tabdedica2dcmri}
            $$
         \begin{array}{l r c c c }
            \hline
            \noalign{\smallskip}
            \hline
            \noalign{\smallskip}
\mathrm{Subclump} & N_{\rm S} & \alpha({\rm J}2000),\,\delta({\rm J}2000)&\rho_{
\rm S}&\chi^2_{\rm S}\\
& & \mathrm{h:m:s,\degree:\arcmm:\arcs}&&\\
         \hline
         \noalign{\smallskip}
\mathrm{RXCJ1825main} & 83&18\ 25\ 24.1+30\ 26\ 16&1.00&25\\
\mathrm{CIZA1824}     & 40&18\ 24\ 02.8+30\ 31\ 04&0.57&15\\
\mathrm{RGregion}     & 42&18\ 24\ 15.2+30\ 23\ 22&0.44&10\\
\mathrm{SGregion}     & 48&18\ 24\ 28.5+30\ 18\ 04&0.37& 9\\
\mathrm{MiddlePeak}   & 25&18\ 24\ 41.2+30\ 29\ 19&0.36& 8\\
\mathrm{RXCJ1825NE}   & 24&18\ 25\ 45.9+30\ 28\ 20&0.32& 8\\
              \noalign{\smallskip}
              \noalign{\smallskip}
            \hline
            \noalign{\smallskip}
            \hline
         \end{array}
$$
         \end{table}

\begin{figure}
\resizebox{\hsize}{!}{\includegraphics{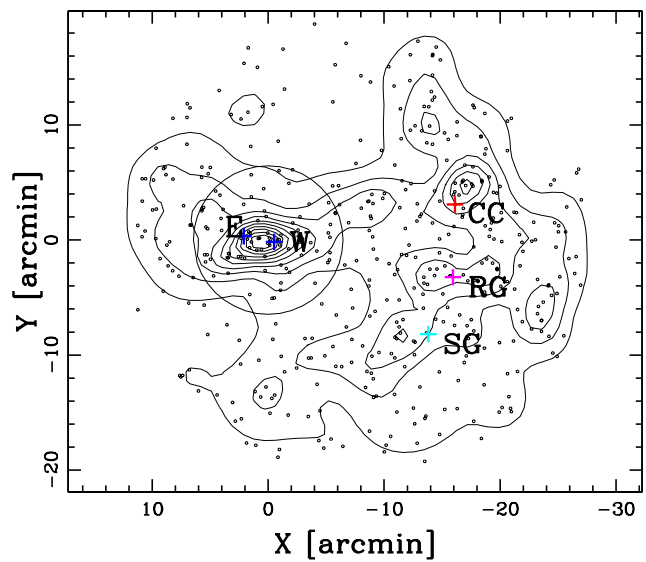}}
\caption
{Spatial distribution on the sky and relative isodensity contour map
  of the 450 photometric members with $r \le 20$, obtained with the
  2D-DEDICA method.
    The X-ray centroid in RXCJ1825 is taken as the center of the whole
  complex.
The position of 
  prominent galaxies (BCGs of RXCJ1825, BCG of CIZAJ1824, SG, RG) are
  indicated by crosses (blue, red, cyan, and magenta, respectively).
  The region of RXCJ1825 within 0.5 \h is enclosed by the circle.
}
\label{figk2cmri}
\end{figure}

Figure~\ref{figk2cmri} shows the contour map for the 450 photometric
members having $r \le 20$, that is $\sim 4$ mags fainter than
$M^*$, the characteristic absolute magnitude of the luminosity
function of galaxies in clusters.  Table~\ref{tabdedica2dcmri} lists
the significant peaks detected in the same region sampled by our
redshift measures.  We confirm the results obtained in the
spectroscopic sample. As above, RXCJ1825 is detected and shows a
structure elongated in the East--West direction, with a small feature
at North--East. CIZAJ1824 is clearly detected, too. The South--West
region of the field is now better sampled and reveals a few
structures.

\subsection{3D structure}
\label{3D}

The existence of correlations between positions and velocities
is a powerful footprint of real substructures.

We analyzed the presence of a velocity gradient performing a multiple
linear regression fit to the observed velocities with respect to the
galaxy positions in the plane of the sky computing the coefficient of
multiple determination ($RC^2$, NAG Fortran Workstation Handbook
\citeyear{nag1986}). The position angle on the celestial sphere of the
velocity gradient is PA$=-78_{-16}^{+14}$ degrees (measured
counter--clock--wise from north), that is higher--velocity galaxies lie
in more western regions, with the velocity gradient pointing from 
RXCJ1825 to CIZAJ1824 as expected.
We followed \citet{girardi1996} to asses the significance of this
velocity gradient and performed 1000 Monte Carlo simulations by
randomly shuffling the galaxy velocities and for each simulation we
determined the coefficient of multiple determination $RC^2$.  The
significance of the velocity gradient is determined as the fraction of
times in which the $RC^2$ of the simulated data is smaller than the
observed $RC^2$.  We found that the velocity gradient is strongly
significant at the $>99.9\%$ c.l..

In order to study the velocity field of the cluster complex, we used
the statistics and the bubble plot of Dressler \& Schectman
(\citealt{dressler1988}, hereafter DS-test).  For each galaxy, the
deviation $\delta$ is defined as $\delta_i^2 =
[(N_{\rm{nn}}+1)/\sigma_{\rm{v}}^2][(\overline {\rm v_l} - \overline
  {\rm v})^2+(\sigma_{\rm v,l} - \sigma_{\rm v})^2]$, where the
subscript ``l'' denotes the local quantities computed over the group
formed of the galaxy itself and its $N_{\rm{nn}}=10$ neighbors.  The
cumulative deviation of the local kinematical parameters (mean
velocity and velocity dispersion) from the global cluster parameters
is given by the value $\Delta$, the sum of the $\delta_i$ of the
individual $N$ galaxies.  The significance of $\Delta$, that is how
far is the system from dynamical equilibrium, is checked by running
1000 Monte Carlo simulations, randomly shuffling the galaxy
velocities.

Girardi et al. (\citeyear{girardi1997a}; \citeyear{girardi2010})
introduced two variations of the DS-test where the contributes of the
local mean and dispersion are considered separately. The kinematical
indicator based on the local mean was proved to be particular useful
since that based on the velocity dispersion needs an elaborate
treatment in presence of very large samples (e.g.,
\citealt{girardi2015}). Following the methodology of
\citet{girardi2010}, we used the kinematical indicator based on the
deviation of the local mean, $\delta_{{\rm v},i}^2= [(N_{\rm
    nn}+1)/\sigma_{\rm v}^2][(\overline {\rm v_l} - \overline {\rm
    v})^2]$, and the significance of this test (DSv-test) was
determined as in the standard DS-test.

\begin{figure}
\centering 
\resizebox{\hsize}{!}{\includegraphics{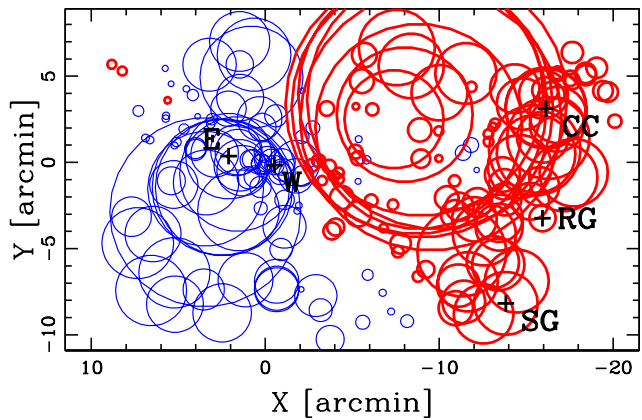}}
\caption
{Spatial distribution of the 198 members of the Lyra complex, each marked
  by a circle: the larger the circle, the larger is the deviation
  $\delta_{{\rm V},i}$ of the local mean velocity from the global mean
  velocity (the so-called bubble plot).  Thin/blue and thick/red
  circles show where the local value is smaller or larger than the
  global value.  The plot is centered on the X-ray centroid of
  RXCJ1825 and prominent galaxies are indicated, for an easier
  comparison with Fig.~\ref{figk2z}.  }
\label{figds}
\end{figure}

  Both the DS- and DSv-tests reveal that the Lyra
  system is not relaxed, at the $>99.9\%$ c.l..  Figure~\ref{figds}
  points out the presence of RXCJ1825 and CIZAJ1824 as low and high
  velocity regions in the velocity field. It also shows that the
  South--West region is characterized by high velocities, comparable to
  those in CIZAJ1824. The intermediate region between RXCJ1825 and
  CIZAJ1824 is characterized by a still higher velocity region. This
  suggests that the MiddlePeak is not a structure connecting the two
  clusters, which instead should be characterized by an intermediate
  velocity.

\subsection{Red galaxies}
\label{red}

In the local Universe red and/or passive galaxies are known to
populate the cluster cores and are used to trace important galaxy
systems or structures (e.g., \citealt{lubin2000};
\citealt{girardi2015}).  We used the $g$--$r$ vs. $r$ CMR to separate
red from blue spectroscopic members. The fitted relation is
$g$--$r$=1.334-0.032$\times r$, based on 116 galaxies. We define blue
galaxies those bluer than 0.15 mag with respect to the typical color
of red sequence galaxies (see Fig.~\ref{figcm}), where the value
  of 0.15 is about two times the error on the intercept. According to
this definition, out of the 194 galaxies with both redshift and
magnitude information, 153 and 41 are defined red and blue galaxies,
respectively.

\begin{table}
  \caption[]{2D substructure from the analysis of  red spectroscopic members.}
         \label{tabdedica2dred}
            $$
         \begin{array}{l r c c c }
            \hline
            \noalign{\smallskip}
            \hline
            \noalign{\smallskip}
\mathrm{Subclump} & N_{\rm S} & \alpha({\rm J}2000),\,\delta({\rm J}2000)&\rho_{
\rm S}&\chi^2_{\rm S}\\
& & \mathrm{h:m:s,\degree:\arcmm:\arcs}&&\\
         \hline
         \noalign{\smallskip}
\mathrm{RXCJ1825main}   & 63&18\ 25\ 24.5+30\ 26\ 15&1.00&21\\
\mathrm{CIZAJ1824}      & 23&18\ 24\ 03.4+30\ 30\ 45&0.43&17\\
\mathrm{RXCJ1825NE}     & 19&18\ 25\ 44.7+30\ 28\ 07&0.35& 9\\
\mathrm{SouthWestPeak}  & 35&18\ 24\ 22.3+30\ 23\ 43&0.30&10\\

              \noalign{\smallskip}
              \noalign{\smallskip}
            \hline
            \noalign{\smallskip}
            \hline
         \end{array}
$$
         \end{table}

\begin{figure}
\centering 
\resizebox{\hsize}{!}{\includegraphics{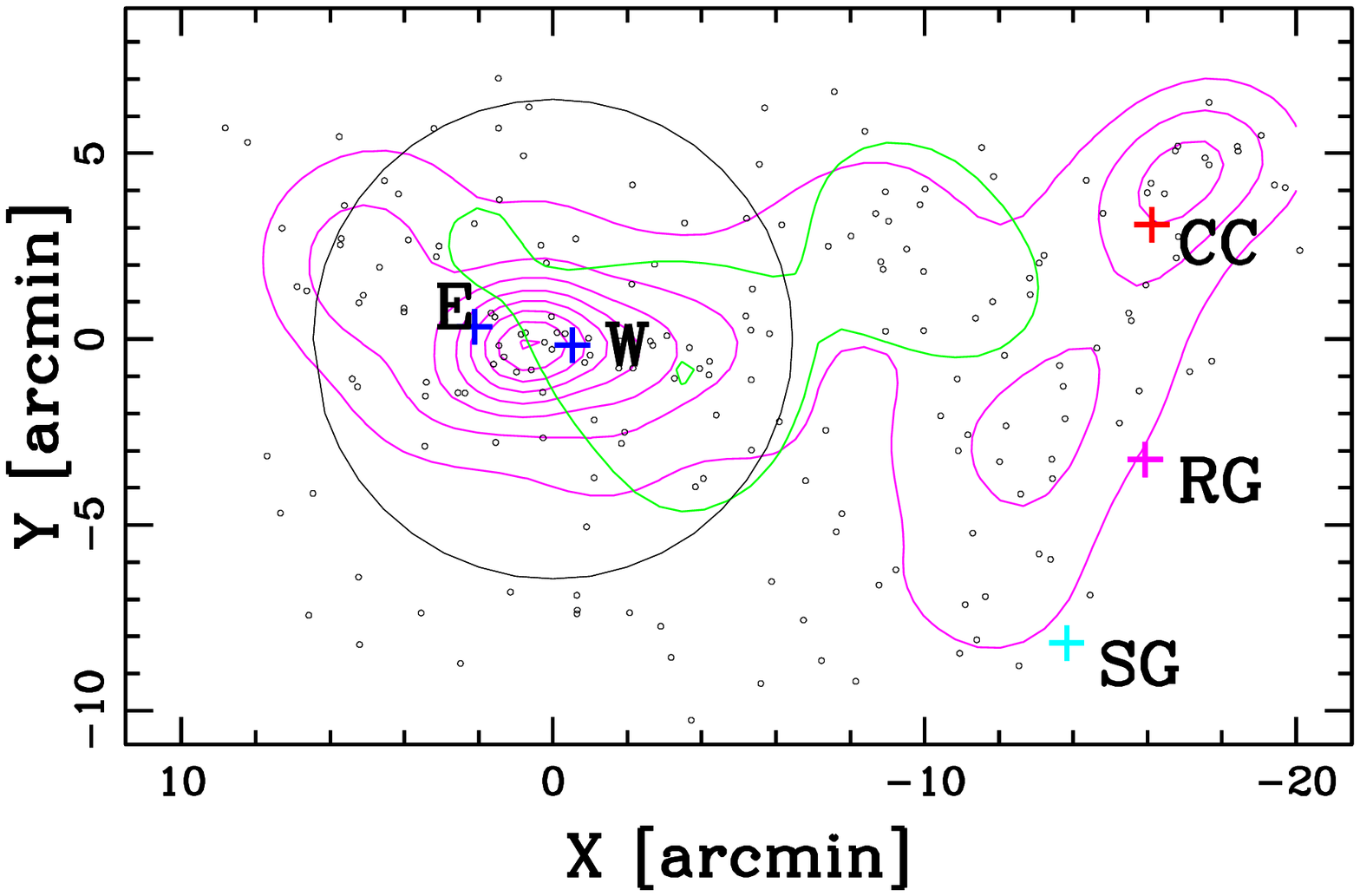}}
\resizebox{\hsize}{!}{\includegraphics{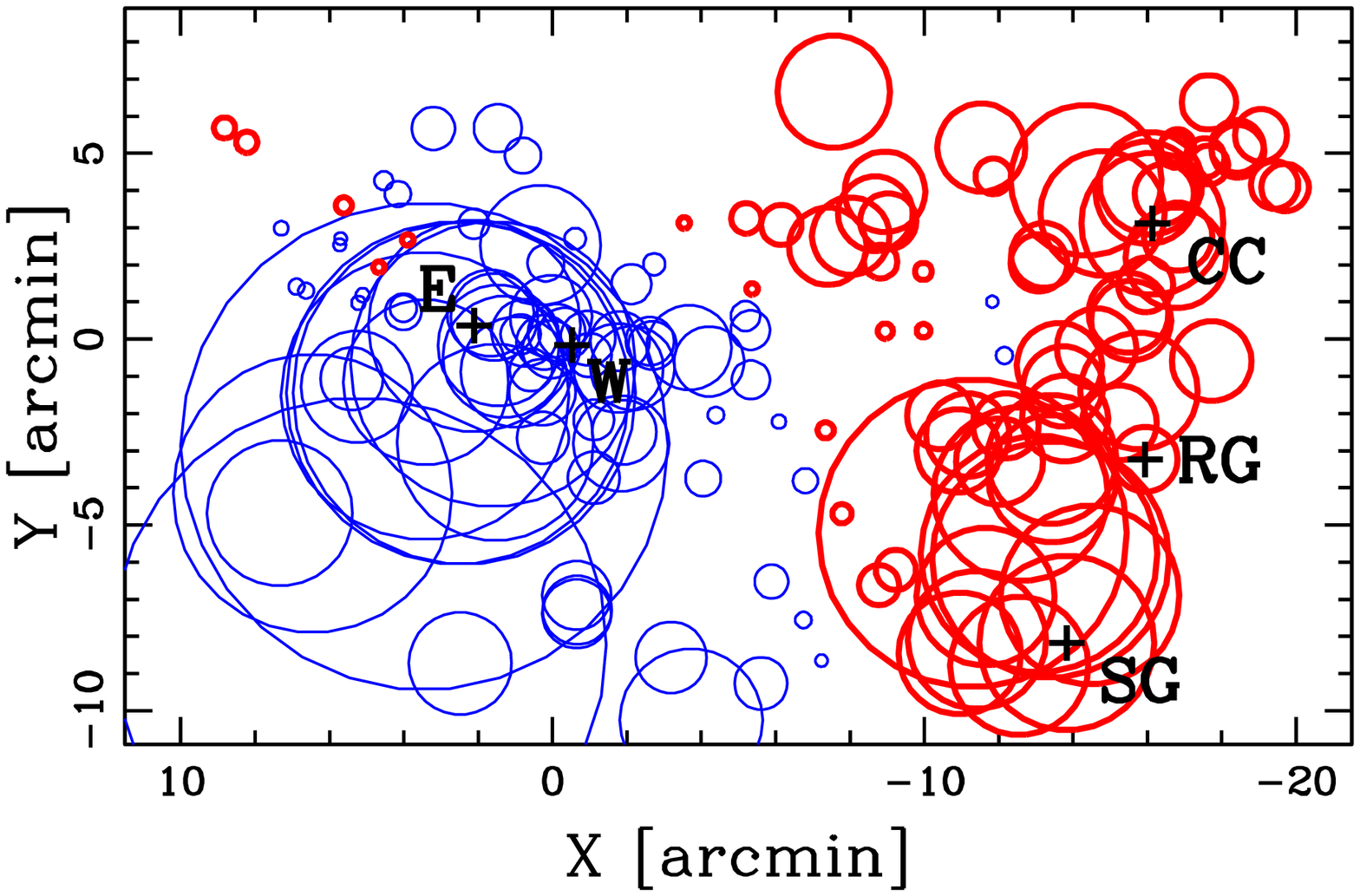}}
\caption
    {{\em Upper panel.} As Fig.~\ref{figk2z}, but with isodensity
      contours for the 41 blue galaxies (in green color) and 153
      red galaxies (in magenta color).  {\em Lower panel.} As
      Fig.~\ref{figds}, but for the 153 red galaxies.}
\label{figred}
\end{figure}

Figure~\ref{figred} (upper panel) shows the isodensity contours
computed for blue and red galaxies separately using the 2D-DEDICA
method. The relevant peaks obtained for red galaxies are listed in
Tab.~\ref{tabdedica2dred}. Figure~\ref{figred} (lower panel) shows the
result of the DSv-test. This analysis based on red galaxies generally
confirms that based on all galaxies.  According to the DS- and
DSv-tests, the presence of substructure is significant at the
$>99.9\%$ c.l. and RXCJ1825 and CIZAJ1824 are always detected as low
and high velocity regions in the velocity field.  The velocity
gradient is very significant ($>99.9\%$ c.l.)  and points from
RXCJ1825 toward CIZAJ1824 (PA$=-78_{-17}^{+14}$ degrees), in agreement
with that found for all galaxies.

The analysis based on red galaxies differs in two points.  A
South--West peak is now detected, although just over the threshold of
$\rho_S=$0.3 (SouthWestPeak in Table~\ref{tabdedica2dred}).  More
interestingly, the MiddlePeak is not longer detected by our 2D-DEDICA
analysis and it seems rather related to the presence of blue galaxies
(see blue contours in Fig.~\ref{figred}, upper panel). The comparison
of Fig.~\ref{figred} (lower panel) with Fig.~\ref{figds} shows that
the velocity peculiarity of the intermediate region between the two
clusters is no longer present.  We conclude that the above detection
of a MiddlePeak is due to a few galaxies, maybe a loose group or a
filament, in the phase of accretion from the field, rather than to an
important structure connecting the two clusters.

\begin{figure}
\centering 
\resizebox{\hsize}{!}{\includegraphics{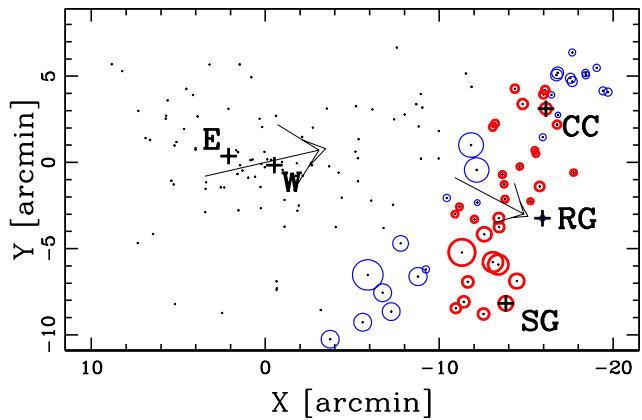}}
\caption
    { Bubble plot for the 58 red galaxies belonging to the West region
      (CIZAJ1824+SouthWestPeak) where the velocity gradient, pointing
      from low to high velocities, is indicated by the little arrow.
      The spatial distribution of all red galaxies is indicated by
      small black dots and the related velocity gradient is indicated
      by the big arrow, pointing from RXCJ1825 to CIZAJ1924.  }
\label{figdsredCCSW}
\end{figure}

The analysis of the sample of red galaxies confirms that the South--West
region of the field is characterized by high velocity galaxies (see
Fig.~\ref{figred}, lower panel). We focused our attention onto the
West region considering only the 58 red galaxies belonging to the
CIZAJ1824 and SouthWestPeak (see Table~\ref{tabdedica2dred}).  Using
the DS- and DSv-tests, we found no evidence for substructure also when
considering smaller numbers of neighbors to compute local quantities
(down to $N_{\rm{nn}}=5$ neighbors) suggesting that CIZAJ1824 and
SouthWestPeak are characterized by very similar kinematical
properties.  The velocity gradient, significant at the 95\% c.l.,
points from North--East to South--West (PA$=-118_{-12}^{+15}$ degrees,
see Fig.~\ref{figdsredCCSW}).  These results are somewhat unexpected.
In fact, one would expect that the South--West region were populated by
galaxies of both the two clusters, especially by those of the rich
RXCJ1825 cluster, and showed an intermediate velocity, with a strong
gradient pointing from South--West to CIZAJ1824.  We conclude that
the Lyra region, and in particular its South--West region, are more complex
than expected (see the below section, too).

\subsection{SG and the SW region}
\label{SW}

\begin{figure}
\centering 
\resizebox{\hsize}{!}{\includegraphics{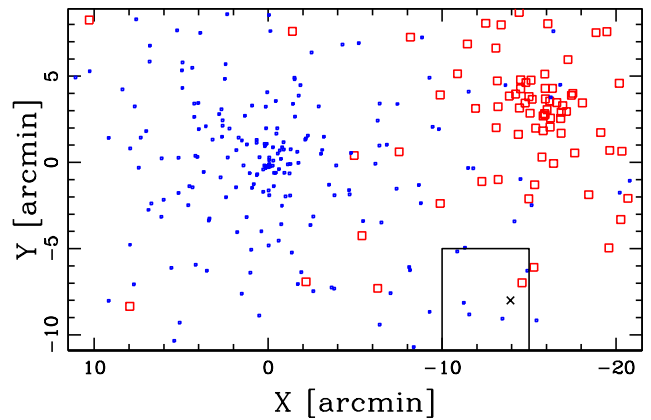}}
\caption
    {Spatial distribution of simulated galaxies of the
      RXCJ1825+CIZAJ1824 system. Blue and red points indicate
      galaxy-points of RXCJ1825 and CIZAJ1824, at low and high
      velocities, respectively. The black cross indicates the position
      of SG and the rectangle delimit a region close to SG, where the
      real velocity field is characterized by high velocities
      (cfr. with Figs.~\ref{figds} and ~\ref{figred}, lower panel).}
\label{figxy2R200min}
\end{figure}

To check the peculiarity of the observed velocity field, we made a
simple Monte Carlo simulation to reproduce the position on the sky of
galaxies of RXCJ1825 and CIZAJ1824.  We used the NFW model for the
galaxy distribution and our mass estimates derived in
Sect.~\ref{mass}. Figure~\ref{figxy2R200min} shows the result of a
simulation where we fixed 220 and 80 galaxy-points for RXCJ1825 and CIZAJ1824
within the respective $R_{200}$ radii\footnote{The radius $R_{\delta}$ is the radius of a sphere
  with mass overdensity $\delta$ times the critical density at the
  redshift of the galaxy system.}, that is a number of galaxies
proportional to the respective $M_{200}$. To be more realistic, we
allowed the simulation to fill with galaxies the whole $2R_{200}$
region, following the same NFW model, for a total number of 327+118
galaxy-points. We do not simulate the Gaussian velocity distribution of each
cluster but galaxy-points of RXCJ1825 and CIZAJ1824 are all assumed to have
the same velocity of the respective parent cluster (see blue and red
points in Fig.~\ref{figxy2R200min}).  The comparison of
Fig.~\ref{figxy2R200min} with Figs.~\ref{figds} and \ref{figred} (lower
panel) show that the simulated field is populated by low velocity
galaxy-points while the real field is populated by high velocity
galaxies. This analysis confirms that the high velocity field detected
in the South--West region is well far from that expected in the context
of a system formed by only two relaxed clusters.

\begin{figure}
\centering 
\resizebox{\hsize}{!}{\includegraphics{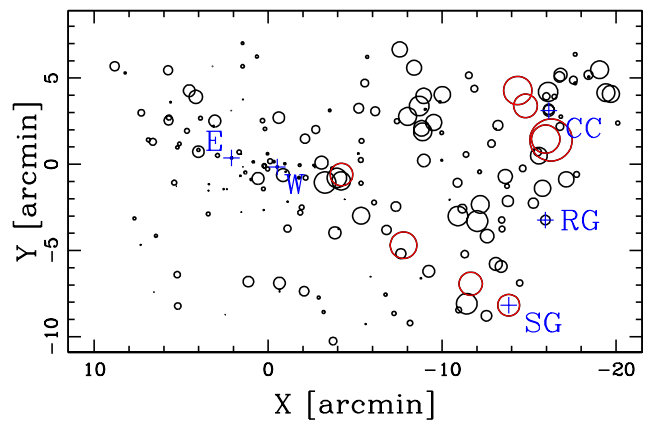}}
\caption
    {Spatial distribution of the 198 galaxies of the Lyra complex,
      each marked by a circle: the larger the circle, the larger is
      the velocity.  The eight galaxies with velocity equal or larger
      than SG are pointed out by red circles. The position of
      prominent galaxies is indicated, too.
    }
\label{figxyhv}
\end{figure}

In the South--West region, the brightest galaxy is SG, which is the
fifth brightest galaxy of the whole Lyra field analyzed here.  In
general, most luminous galaxies in clusters are related to groups as
shown when accurate dynamical and/or gravitational lensing analyses
are performed (e.g., \citealt{girardi2008} and refs. therein).  SG has
a quite high velocity, more than 1000 \ks higher than that of
BCG-CC. In order to detect galaxies related to SG minimizing possible
contamination from the two clusters, we selected galaxies having
velocities larger than SG. Fig.~\ref{figxyhv} shows the positions of
these galaxies in the sky. Four of these galaxies are close to the
center of CIZAJ1824 and might be the high tail of the velocity
distribution of its galaxy population.  The other three are aligned
between SG and RXCJ1825 along a stretched structure.

\section{Disentangling RXCJ1825 from CIZAJ1824}
\label{dise}

Our analysis of the 2D galaxy distribution of the Lyra complex well
detects the two individual clusters and indicates that RXCJ1825 is
more massive than CIZAJ1824 by a factor 2-4 depending if we consider the
density of the peaks or the richness of the corresponding samples, and
the RXCJ1825main+NE peaks vs. the CIZAJ1824 sample or only
RXCJ1825main vs. the CIZAJ1824 (see Tabs.~\ref{tabdedica2d},
~\ref{tabdedica2dcmri}, and ~\ref{tabdedica2dred}).

However the velocity distribution does not show evidence for the
presence of two separated peaks (see Fig.~\ref{fighisto}, inset)
and both 1D-DEDICA and 1D-KMM method (\citealt{ashman1994}) failed in
finding a bimodality.  Of consequence we consider the values of the
velocities of the BCGs, as measured from redshift, fair signposts of
the velocities of the two clusters. This is likely true for CIZAJ1824,
a very relaxed cluster according to Clavico et
al. (\citeyear{clavico2019}), for which we measure ${\rm v}_{{\rm
    BCG}-{\rm CC}}=21215\pm50$ \kss, and also for RXCJ1825 where the
two BCGs have the same velocity within the errors with an average
value of ${\rm v}_{{\rm BCG}-{\rm EW}}=19340\pm67$ \kss.

Analyzing the two samples corresponding to the peaks CIZAJ1824 and
RXCJ1825main detected in the 2D-DEDICA analysis of the galaxy
  distribution, we obtained $\left<\rm{v}\right>=21102\pm 214$ \ks
and $\left<\rm{v}\right>=19618\pm 152$ \kss, in agreement within the
errors with the measured velocities of the BCGs. For both samples
we obtained quite large estimates of velocity dispersion, $\sigma_{\rm
  v}=1089_{-115}^{+178}$ \ks and $\sigma_{\rm v}=1254_{-110}^{+144}$
\kss.  However we expect that each sample may be contaminated by some
galaxies belonging to the companion system or other interveining
substructures due to the projection effects. We stress that: i)
CIZAJ1824, the poorer cluster, is expected to be the more
contaminated; ii) the contamination is expected to be larger among
blue galaxies due to the field or the outskirts of the companion
cluster; iii) the $\sigma_{\rm v}$ estimator is much less robust than
$\left<\rm{v}\right>$ to the inclusion of interloper galaxies, the
effect of inclusion is generally an increasing of $\sigma_{\rm v}$
estimates.

\begin{figure}
\centering 
\resizebox{\hsize}{!}{\includegraphics{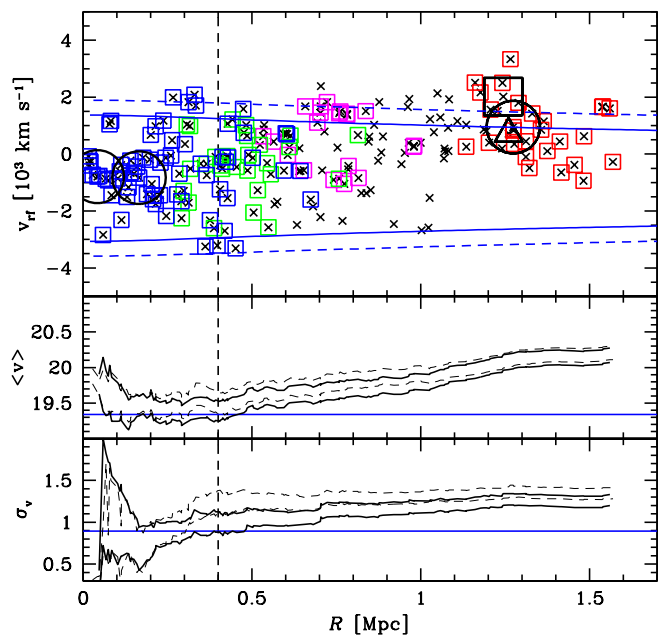}}
\caption
{{\em Top panel:} System rest--frame velocity vs. projected
  clustercentric distance for the 198 galaxies in the whole cluster
  complex. Blue, green, magenta, and red squares indicate galaxies
  belonging to the four peaks detected by the 2D-DEDICA method, that
  is RXCJ1825main, RXCJ1825NE, MiddlePeak, CIZAJ1824 (see
  Fig.~\ref{figk2z} and Tab.~\ref{tabdedica2d}). The two close large
  circles indicate the two BCGs of RXCJ1825, close to the cluster
  center, here assumed the X-ray centroid of RXCJ1825. The other large
  circle indicates BCG-CC of the CIZAJ1824 cluster.  The large square
  indicates the SG galaxy and the triangle indicates the RG
  radio-galaxy.  Solid blue curves show the limits due to the escape
  velocity in RXCJ1825 assuming our estimate for its mass of
  (as based on $\sigma_{\rm v}$ estimated using red galaxies, see
  Sect.~\ref{mass}). The dashed curves refer to $\sigma_{\rm v}$
  estimated using all galaxies.  {\em Middle and bottom panels:}
  integral profiles of mean velocity and velocity dispersion (only the
  one-$\sigma$ error bands are shown) which, by definition, converge
  to the global values of the whole complex.  Solid and dashed lines
  refer to red and all  galaxy populations. The vertical axes are in
  units of $10^3$ \kss.  In the {\em middle panel} the blue horizontal
  line indicates the mean velocity of the two BCGs of RXCJ1825. In the
  {\em bottom panel}, the blue horizontal line indicates the value of
  the X-ray temperature of RXCJ1825 estimated by Clavico et al. (\citeyear{clavico2019})
  and here transformed in $\sigma_{\rm V}$ assuming the density-energy
  equipartition between ICM and galaxies, i.e.  $\beta_{\rm spec}=1$
  (see Sect.~\ref{discu}).  In the three panels the vertical dashed
  line contains the central region of RXCJ1825 suggested to be uncontaminated by
  galaxies of the companion cluster or other substructures.  }
\label{figprof}
\end{figure}

In order to disentangle the two systems and obtain reliable estimates
for the velocity dispersion, we performed the analysis of the mean
velocity profiles and velocity dispersion profiles (e.g.,
\citealt{girardi2016}).  The top panel of Fig.~\ref{figprof} shows the
complexity of the distribution of galaxies of the Lyra complex in the
projected phase-space, that is the rest-frame velocity ${\rm v}_{\rm
  {rf}}=({\rm v}-\left<\rm{v}\right>)/(1+z)$ vs. the clustercentric
radius $R$. As for the system redshift $z$ (and the related
$\left<\rm{v}\right>=cz$), we used that of the full Lyra system which
is assumed to fix the cosmological distance of the system and thus the
\h scale for $R$. The following analysis is independent of this
assumption. As for the center, the X-ray centroid of RXCJ1825 is
assumed.

Figure~\ref{figprof} (middle panel) presents the integral mean
velocity profile for all member galaxies and for red galaxies only.
It is shown that $\left<\rm{v}\right>$ agrees with ${\rm v}_{{\rm
    BCG}-{\rm EW}}$ in the central region of RXCJ1825 and that the
inclusion of more galaxies at larger clustercentric distances causes
an increasing of the value of the mean velocity, likely due to the
contamination of galaxies belonging to CIZAJ1824 or other minor
substructures, up to reach the global value of $\left<\rm{v}\right>$.
The inspection of the mean velocity profile points out that
  $\left<\rm{v}\right>$ starts to increase at about 0.4 \hh, with
  $\left<\rm{v}\right>$ being already 1-sigma higher than the 
  mean velocity of the two BCGs of RXCJ1825 at 0.5 \hh.  Therefore, we can
assume that the contamination is not relevant for $R<0.4$ \hh.

Figure~\ref{figprof} (bottom panel) presents the integral
velocity-dispersion profile (hereafter VDP) for all member galaxies
and for red galaxies only.  The VDP of relaxed clusters is expected to
have a gentle decline down to the global value of the velocity
dispersion while the contamination of a companion cluster produce a
sharp increase (e.g., cfr. Fig.~2 of \citealt{girardi1998} with Fig.~2
of \citealt{girardi1996}).  As for RXCJ1825, its VDP is sharply
declining down to $\sim$ 0.15 Mpc (but with large uncertainties) and
then increasing, very slightly in the case of the red galaxies. Since
Fig.~\ref{figprof} (top panel) shows that the galaxies of the
CIZAJ1824 peak are still well far, we assumed the velocity dispersion
computed using galaxies within $R< 0.4$ \h is a reliable estimate of
the velocity dispersion of the RXCJ1825 galaxy population. We computed
$\sigma_{\rm v}=1244_{-131}^{+133}$ \ks using all 62 galaxies and
$\sigma_{\rm v}=995_{-125}^{+131}$ \ks using the 49 red
galaxies. Since the red population is likely less affected by
contamination, to be more conservative we adopt the latter value as
the estimate of the velocity dispersion in RXCJ1825.  The minimum
point of VDP, $\sigma_{\rm v}=628_{-200}^{+313}$ \kss, is found at
$R=0.166$ \hh, which is indeed comparable to the standard size of
cluster cores when the galaxy profile is fitted with King-like models
(see \citealt{girardi1995} and refs. therein).

\begin{figure}
\centering 
\resizebox{\hsize}{!}{\includegraphics{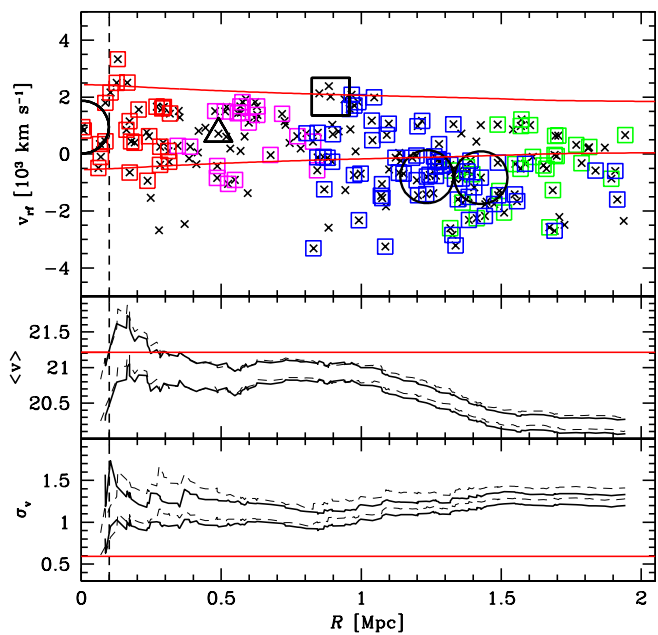}}
\caption
    {As Fig.~\ref{figprof} but centered on BCG-CC, the adopted center
      of CIZAJ1824. Dashed red curves in the {\em top panel} show the
      limits due to the escape velocity in CIZAJ1824 assuming our
      estimate for its mass.  In the {\em middle
        panel} the red horizontal line indicates the velocity of
      BCG-CC. In the {\em bottom panel}, the red horizontal line
      indicates the value of the X-ray temperature of CIZAJ1824
      estimated by Clavico et al. (\citeyear{clavico2019}).
      In the three panels the vertical dashed
      line contains the very central region of CIZAJ1824
discussed in the text.
}
\label{figprofCC}
\end{figure}

The results of the same analysis but for CIZAJ1824 is shown in
Fig.~\ref{figprofCC}, where the center is fixed on BCG-CC.  Following
the above approach, we note that $\left<\rm{v}\right>$ shows a
decreasing already in the very central regions thus we suspect that
the contamination by RXCJ1825 or other substructures is already
important out of $\sim 0.1$ \h (0.15 \hh).  The velocity dispersion
computed in this small region is $\sigma_{\rm v}=858_{-226}^{+330}$
\ks ($\sigma_{\rm v}=785_{-161}^{+264}$ \kss), based on six (seven)
galaxies.  However this estimate of $\sigma_{\rm v}$ is not supported
by the analysis of the VDP.

\begin{figure*}[!ht]
\centering 
\includegraphics[width=9cm,angle=90]{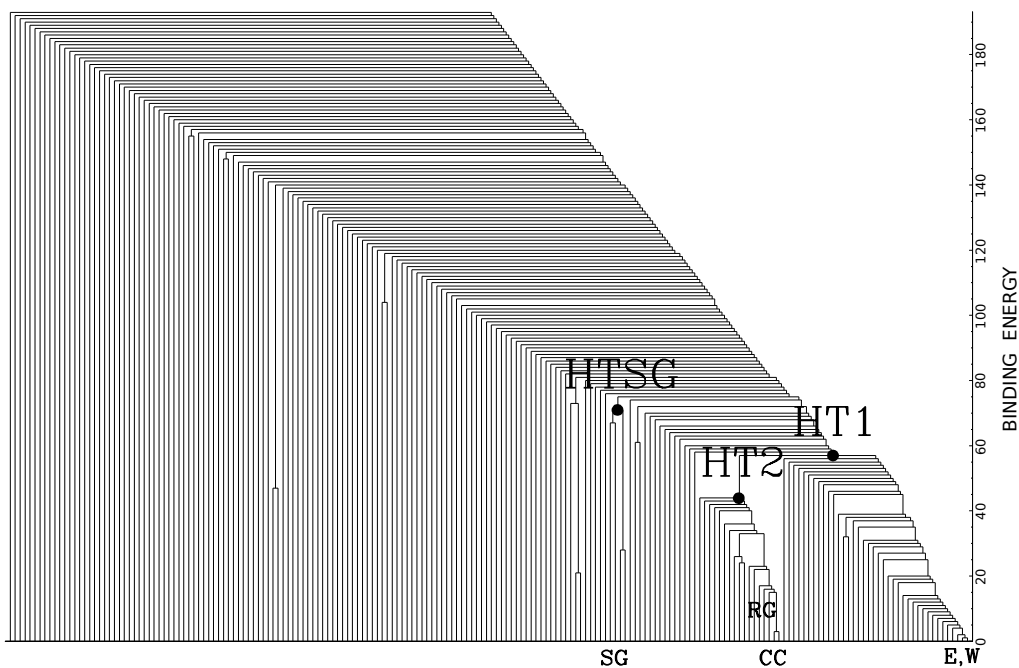}
\caption{ Dendogram obtained through the Serna \& Gerbal algorithm
  applied to the 194 members with available magnitudes (here the case
  for $M/L_r$=150 \mll).  The y-axis indicates the binding energy, here in
  arbitrary units, with the deepest negative energy levels on the
  bottom. The position of the various galaxies are shown along the
  x-axis, where small labels indicate prominent galaxies. Big labels
  indicate nodes of the structures discussed in the text.}
\label{figHT}
\end{figure*}

\begin{figure}
\centering
\resizebox{\hsize}{!}{\includegraphics{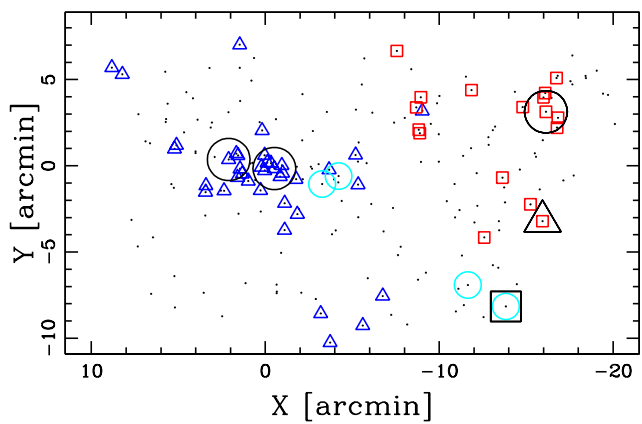}}
\caption
    {Spatial
      distribution on the sky of 194 galaxies of the Lyra complex  having available
      magnitudes.  Blue triangles, red squares, and cyan circles
      indicate galaxies of HT1, HT2, and HTSG as obtained from
      Serna \& Gerbal method (see Fig.~\ref{figHT}). Large black symbols
      indicate BCGs (circles), SG (square), RG (triangle).
    }
\label{figshowgerbal}
\end{figure}

We resorted to using the method devised by \citet{serna1996}, know as
H-tree method or Serna-Gerbal method (e.g., \citealt{durret2010};
\citealt{adami2018}). This method uses a hierarchical clustering
analysis to determine the relationship between galaxies according to
their relative binding energies. The method assumes by definition that
the redshift difference are of kinematical nature. The method also
assumes a constant value for the mass-to-light ratio of galaxies and
Serna \& Gerbal suggested a value comparable to that of clusters. Out
of our catalog of 198 member galaxies we use the 194 galaxies having
available magnitudes. We considered values of $M/L_r$=100, 150, and
200 \ml as suggested by large statistical studies (e.g.,
\citealt{girardi2000}; \citealt{popesso2005}; \citealt{proctor2015}).
The (gross) results are quite robust against the choice of the value
of $M/L_r$. Figure~\ref{figHT} shows the resulting dendogram for
$M/L_r$=150 \mll, where the total energy appears vertically.  At the
deepest potential hole there are BCG-E and BCG-W of RXCJ1825.  There
is no relevant substructure in the most part of the cluster down to a
level where a group hosting BCG-CC (hereafter HT2) departs from the
main tree which continues with the group hosting BCG-E and BCG-W
(hereafter HT1), as shown in Fig.~\ref{figHT}.  The groups HT1 and HT2
are formed of 42 and 19 galaxies, respectively.  The mean velocities
of HT1 and HT2 are consistent with the mean velocities of the two
clusters recovered with other methods (see above). The spatial
position of the galaxies of HT1 and HT2 are also shown in
Fig.~\ref{figshowgerbal}.  A similar result is obtained considering
$M/L_r$=100 and 200 \mll.  We conclude that the two clusters are the
only important structures detected in the field.

  \begin{table*}
  \caption[]{Properties of RXCJ1825 and CIZAJ1824 clusters.}
         \label{tabv}
            $$
         \begin{array}{l c c c c c c}
            \hline
            \noalign{\smallskip}
            \hline
            \noalign{\smallskip}
 \mathrm{Cluster}& \mathrm{Center}&z&\rm{v}&\sigma_{\rm V}\;\;\;& R_{200}&M_{200}\\
  & \mathrm{\alpha},\mathrm{\delta}\,(\mathrm{J}2000)&&\mathrm{km\ s^{-1}}\;\;& \mathrm{km\ s^{-1}}\;\;\; & \mathrm{Mpc}& 10^{15}\mathrm{M}_{\sun}\\

 \mathrm{RXCJ1825}^{a}&
 18^{\mathrm{h}}25^{\mathrm{m}}21\dotsec77,+30\degree 26\arcmm
 25.3\arcs &0.0645\pm0.0002&19340\pm67& 995_{-125}^{+131}& 2.1\pm0.2&1.1\pm0.4\\

  \mathrm{CIZAJ1824}^{b}&18^{\mathrm{h}}24^{\mathrm{m}}06\dotsec98,
  +30\degree 29\arcmm 30.4\arcs&0.0708\pm0.0002&21215\pm60&700\pm50& 1.4\pm0.1& 0.4\pm 0.1\\

  \noalign{\smallskip}
             \noalign{\smallskip}
            \hline
            \noalign{\smallskip}
            \hline
         \end{array}
$$
         
\begin{list}{}{}  
\item[$^{\mathrm{a}}$] The center is the X-ray centroid (Clavico et al. 2019).
\item[$^{\mathrm{a}}$] The center is the position of BCG-CC.
\end{list}
         \end{table*}

Depending on the adopted value of $M/L_r$, the number of members of
HT2 ranges in the 17--19 interval and the value of $\sigma_{\rm v}$ in
the 678--743 \ks interval.  We decided to adopt the average value of
$\sigma_{\rm v}=700\pm50$ \kss, as a reliable estimate of the velocity
dispersion of CIZAJ1824 galaxies.  As for HT1, the number of members
ranges in the 35--49 interval and the value of $\sigma_{\rm v}$ in the
591--766 \ks interval. The average value is $\sigma_{\rm v}=650\pm100$ \kss,
which is indeed much smaller than that obtained from our VDP analysis
for RXCJ1825, but agrees with that measured for the core within $\sim
0.166$ \hh. We conclude that the HT1 group should be interpreted as
the core -- plus galaxies strictly bound to it -- of RXCJ1825.

We also applied the Serna \& Gerbal method to the sample of red
galaxies with the addition of BCG-W. In fact, although BCG-W is
slightly bluer than our definition of red galaxies, it is an
elliptical galaxy and cannot be neglected in this method where
galaxies are weighted with their luminosities.  Very reassuringly, the
results on this sample of 154 galaxies is comparable to those obtained
for the full sample, with the detection of HT1 and HT2 having
$\sigma_{\rm v}$ values in the above ranges.

When using all galaxies, the SG galaxy is assigned to the global
cluster (in the $M/L_r=200$ \ml case) or assigned to a small group of
4 galaxies (HTSG, in the $M/L_r=100$ and 150 \ml cases).  When using
red galaxies, the SG galaxy is always assigned to the global
cluster. Thus we cannot be confident in the detection of a group
related to SG, although we indicate it in Fig.~\ref{figshowgerbal}
where it appears as a very stretched structure.  Finally, the RG
galaxy is assigned only in one case to the HT2 group, otherwise is
part of the global cluster or very poor groups at higher energy level.

Table~\ref{tabv} lists our best estimates of cluster velocity and
velocity dispersion, $\rm{v}$ and $\sigma_{\rm v}$, for RXCJ1825 and
CIZAJ1824, separately.  The listed uncertainties are strictly related
to the method of analysis.

\section{Mass estimates of RXCJ1825 and CIZAJ1824}
\label{mass}

To compute the mass of the two clusters we used the values estimated
for $\sigma_{\rm v}$ in the previous section (see Tab.~\ref{tabv}) and
applied the relation by Munari et al. (\citeyear{munari2013}, their
eq.~1):
\begin{equation}
M_{200}/10^{15}{\rm M}_{\sun}=[\sigma_{\rm v}/A_{\rm 1D}]^{1/\alpha}/h(z),
\end{equation}
where $A_{\rm 1D}=1090$ \kss, the average of the values they proposed,
$1/\alpha=3$, and $h(z)$ is computed using the adopted cosmology of this study.

As for RXCJ1825, we estimated a mass $M_{\rm 200}=1.1\pm 0.4$ \mqui
within $R_{\rm 200}=2.1\pm 0.2$ \hh, where uncertainties of 10\% and
30\% on $R_{200}$ and $M_{200}$ are due to the propagation of the
uncertainty on $\sigma_{\rm v}$.  An additional 10\% of uncertainty on
mass is also added due to the scatter around the theoretical
relation. We used the recipe of \citet{denHartog1996} with the
assumption of a NFW mass density profile (\citealt{navarro1997};
\citealt{Dolag2004}) to derive from our mass estimate the
``caustics'', that is the curves delimiting the region where the
rest-frame velocity ${\rm v}_{\rm rf}$ is smaller than the escape
velocity. The inspection of Fig.~\ref{figprof} (top panel, solid blue
curves) suggests that our mass estimate is adequate to describe the
position of the RXCJ1825 galaxies in the phase-space. In the same
figure we also plot the caustics derived from the mass $M_{\rm
  200}=2.1$ \mqui computed using $\sigma_{\rm v}$=1244 \ks obtained
for all galaxies instead for red ones, which give a kind of external
envelope.  As for CIZAJ1824, we estimated a mass $M_{\rm 200}=3.7\pm
1.1$ \mqua within $R_{\rm 200}=1.4\pm 0.1$ \hh, the caustics to verify
this mass value are plotted in Fig.~\ref{figprofCC} (top panel, solid
red curves).

\section{A bimodal model}
\label{bim}

If we assume that the two clusters are at the same distance from us,
that is their redshift difference is of kinematical nature, the
rest-frame velocity difference between the two cluster is $V=\Delta {\rm
  v}_{\rm rf}=\Delta {\rm v}/(1+z)= 1757$ \kss, and the projected
distance is $D=1.272$ \hh.

  \begin{table}
  \caption[]{Properties of the whole system.}
         \label{tabsys}
            $$
         \begin{array}{c c c c c}
            \hline
            \noalign{\smallskip}
            \hline
            \noalign{\smallskip}
z& \sigma_{\rm V}\;\;\;&M_{\rm sys}&M_{\rm vir,obs.region}&M_{\rm sys,200}\\
 & \mathrm{km\ s^{-1}}\;\;\; &10^{15}\mathrm{M}_{\sun}&10^{15}\mathrm{M}_{\sun}&10^{15}\mathrm{M}_{\sun}\\
0.0674& 1342_{-68}^{+60}& 1.5-3& 1.4& 2.6\pm0.6\\
  \noalign{\smallskip}
             \noalign{\smallskip}
            \hline
            \noalign{\smallskip}
            \hline
         \end{array}
         $$
         \end{table}

\begin{table}
  \caption[]{Parameters of the    bimodal model.}
         \label{tabbim}
            $$
         \begin{array}{c c c}
            \hline
            \noalign{\smallskip}
            \hline
            \noalign{\smallskip}
M_{\rm sys}& V & D\\
10^{15}\mathrm{M}_{\sun}&\mathrm{km\ s^{-1}}\;\;\; &\mathrm{Mpc} \\
1.5-3&1757\pm96& 1.272\\
  \noalign{\smallskip}
             \noalign{\smallskip}
            \hline
            \noalign{\smallskip}
            \hline
         \end{array}
$$         
         \end{table}

To perform the dynamical analysis of the system, one needs a third
observational parameter, the mass of the whole system, $M_{\rm sys}$.
We estimated $M_{\rm sys}$ adding the mass of the two cluster masses.
Since it is known that cluster galaxy density profiles, and likely the
DM profile, extend well out of $R_{200}$, to $2R_{200}$ and outer
(\citealt{biviano2003}; \citealt{rines2013}), the involved global mass
of each system may be assumed to be higher by a factor two.  The mass
of the whole system is then estimated to be in the $M_{\rm sys}=$1.5-3
\mqui range.

As an alternative estimate of the mass of the whole system, assuming
that the system is bound (see below), we computed the mass estimate
based on the virial theorem (\citealt{limber1960}).  Following the
  recipe of \citet{girardi1998}:
\begin{equation}
M_{\rm vir}=3 \pi /2 \times  \sigma_{\rm v} \times R_{\rm v} /G,
\end{equation}
  \noindent where $R_{\rm V}=N^2\sum_{\rm i<j}r_{ij}^{-1}$ depends on
  the projected distance $r_{ij}$ between any pair of the $N$
  galaxies. In principle, this method may overestimate by a factor two
  the mass of a system which is bound but not virialized, but
  numerical simulations show the virial mass estimate to be quite
  accurate also for superclusters near turnaround
  (\citealt{small1998}).  We obtained a mass of $M_{\rm
    vir,obs.region}=1.4$ \mqui for the sampled region, that is a
  projected region of $\sim 2.3$Mpc$\times1.5$ Mpc. Since this region is
  much smaller than the $R_{200}$ region of RXCJ1825, our simple
  computation confirms that we are looking at a very massive system.

  As a third approach, on the assumption that the velocity dispersion
  might be an acceptable proxy for the mass also before the
  virialization, we applied the formula of \citet{munari2013} to the
  velocity dispersion for the whole Lyra complex, $\sigma_{\rm
    v}=1342_{-68}^{+60}$ \kss, and obtained $M_{\rm
    sys,200}$=2.6$\pm0.6$ \mquii, which we interpret as the $M_{200}$
  mass of the cluster which will be formed in the future by the merger
  of RXCJ1825 and CIZAJ1824. This mass value lies in the 1.5-3 \mqui
  range of the above estimated $M_{\rm sys}$. Table~\ref{tabsys}
  summarizes the mass values estimated for the whole system.

To check whether the Lyra system is bound, we computed the two-body
Newtonian criterion for gravitational binding that is stated in terms
of the observables to be
\begin{equation}
V^2D<2GM_{\rm
    sys}\rm{sin}^2\alpha\,\rm{cos}\alpha,
\end{equation}
\noindent where $\alpha$ is the projection angle between the plane of
the sky and the line connecting the centers of the two clumps and $V$
and $D$ are the velocity difference and the projected distance between
the two clusters as above reported.  Eq.~3 is valid in
the case of a pure radial motion.  When adding an orbital component in
addition to the radial component, the true formulation is rather
$V^2D<2GM_{\rm sys}\rm{sin}^2\alpha_V\,\rm{cos}\alpha_R$, that is
allowing $\alpha_V \neq \alpha_R$.  Assuming $M_{\rm sys}=1.5$ \mquii,
the probability that the system is bound is 31\% in the case of radial
orbits and 34\% in the more general case, following \citet{beers1982}
and \citet{hughes1995} for the computation.  Assuming the more
realistic value, $M_{\rm sys}=3$ \mquii, the probability that the
system is bound is $58\%$ and $53\%$ in the two cases.

The above probabilities are estimated from the solid angles without
regard to other constraints. Indeed, redshift surveys of clusters of
galaxies limit cluster-cluster peculiar velocities to $<2000$ \ks
(\citealt{bahcall1986}) and basic arguments indicate that typical
velocities involved in cluster mergers are $\sim 3000$ \ks
(\citealt{sarazin2002}). In fact, the highest values for impact
velocity reported in the literature are of the order of 4000 \ks and
are related to important mergers, very close to the core--core passage
(e.g., the bullett cluster of \citealt{markevitch2002}; Abell 2744 of
\citealt{boschin2006}, see also the discussion in
\citealt{molnar2013}). When values of $\alpha<15$\degree, which lead
to deprojected/real velocities $>6500$ \ks are excluded, the bound
probability is enhanced to $78\%$.

In order to analyze the interaction between the two clusters, we
applied the analytical two--body model introduced by Beers et
al. (\citeyear{beers1982}) and Gregory \& Thompson
(\citeyear{gregory1984}), following the methodology of
\citet{girardi2008}.  This model assumes radial orbits for the clumps
with no shear or net rotation of the system, as in Eq.~3.
Furthermore, the clumps are assumed to start their evolution at time
$t_0=0$ -- here the time 0 of the universe -- with separation $D_0=0$,
and are moving apart or coming together for the first time in their
history. We are assuming that we are seeing the cluster prior to
merging at the time t=12.603 Gyr, the age of the universe at the
redshift of the Lyra system.  The bimodal model solution gives the
total system mass $M_{\rm sys}$ as a function of $\alpha$.

\begin{figure}
\centering
\resizebox{\hsize}{!}{\includegraphics{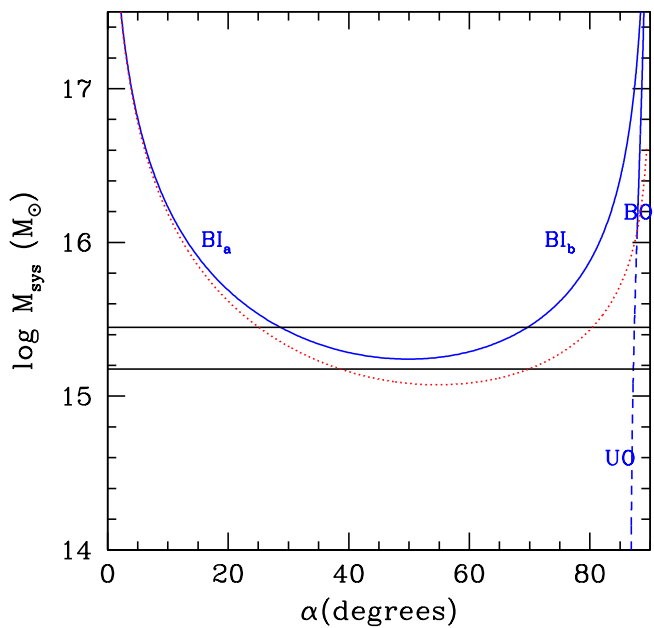}}
\caption
{System mass vs. projection angle for bound and unbound solutions
  (thick solid and thick dashed blue curves, respectively) of the two--body
  model applied to the two clusters.  Labels BI$_{\rm a}$ and
  BI$_{\rm b}$ indicate the bound and incoming, i.e., collapsing
  solutions. Label BO and UO indicate the bound outgoing, i.e.,
  expanding solutions and unbound outgoing solutions. The horizontal
  band gives the range of the observational values of the mass system.
  The thin red dotted curve separates bound and unbound regions
  according to the Newtonian criterion (above and below the thin
  dashed curve, respectively).}
\label{figbim}
\end{figure}

Figure~\ref{figbim} shows the bimodal-model solution in comparison
with our estimate of the mass of the system, which is the most
uncertain observational parameter, here assumed to be in the 1.5--3.0
\mqui range.  The present bound outgoing solutions (i.e. expanding),
BO, is clearly inconsistent with the observed mass.  The
unbound-outgoing solution, UO, is acceptable but only for a
very small range of values of $\alpha$, that is is formally quite
improbable.  Bound ingoing solution solutions (BI) are quite
acceptable and require intermediate values of $\alpha$, in the range
of 30\degree-70\degree. Therefore, the geometry of the merger is that
CIZAJ1824 is in front of RXCJ1825 and moving towards it. The
specific case  with $\alpha=50$\degree implies a real distance of 2 \h and
a real/deprojected velocity of 2300 \kss.  A distance of 2 \h means
that the center of CIZAJ1824 is at a distance of $\sim R_{200}$ from
the center of RXCJ1825, and that the two $R_{500}$-regions of the two
clusters are just in touch but not compenetrated.

We are aware that there are several limitations in characterizing the
dynamics using the bimodal model. The obvious limit is the assumption
that the two systems move on a radial orbit. However, basic arguments
indicate that typical impact parameters should be small, of the order
of 160 kpc (\citealt{sarazin2002}). More important, the model does not
take into account the mass distribution in the two clumps, which
starts to be important when their separation is comparable with their
size. However, in the specific case, the two systems are likely just
approaching, well far from the core-core passage.

\section{Discussion}
\label{discu}

We discuss here our results and some points of our analysis with
particular attention to results based on X-ray and radio data
(\citealt{botteon2019}; \citealt{clavico2019}).

\subsection{Global dynamics}
\label{discubim}

As for our dynamical analysis of Sect.~\ref{bim}, the analytical
two-body model requires that we are looking at the two clusters before
any core-core passage.  This assumption is based on the fact that
\citet{clavico2019} find no excess in the X-ray surface brightness
profile between RXCJ1825 and CIZAJ1824 and the absence of a disturbed
morphology in the X-ray surface brightness of CIZAJ1824.  We also
stress the point that our dynamical analysis implicitly assumed that
the redshift difference between RXCJ1825 and CIZAJ1824 is due to
kinematics, that is both clusters are at the same distance from us.
Instead, in the case the redshift difference is interpreted as a
cosmological distance, the distance between the two clusters is $\sim
30$ \hh.  With the optical data we cannot appreciate this difference,
which implies a difference in optical luminosity of $\sim 20\%$, in
mag of $\sim 0.2$, and not appreciable difference in color (e.g.,
\citealt{lopes2007}). Indeed, the appearance of BCG-CC on the image is
also similar to that of the two RXCJ1825 BCGs.  Using the number
density of galaxy clusters with masses higher than 4\mqua (e.g.,
\citealt{eke1996}; \citealt{vikhlinin2009}), one expects fewer than
$7\times10^{-4}$ clusters as massive as CIZAJ1824 in the sampled
volume, thus the casual presence of two unrelated clusters is quite
improbable.

\subsection{RXCJ1825 and CIZAJ1824}
\label{discucl}

As for RXCJ1825, the two dominant galaxies, BCG-E and BCG-W, have
similar magnitudes, $\Delta m_{12} \sim 0.5$ mag, and are close in
projected position, within $\sim 0.2$ \hh, and in line--of --sight
velocity, since redshift measures are equal within the
uncertainties. The presence of two dominant galaxies in the cluster is
generally taken as evidence of a merger in the past since the giant
ellipticals observed in the center of the cluster are suspected to be
the BCGs of the two previously colliding subclusters.  Indeed,
considering more and more data can make sometimes possible the
detection of the related subclusters (e.g., the case of Coma cluster,
\citealt{colless1996}).

In agreement with the above scenario, we find that the isodensity
contours of the galaxy distribution in the central part of RXCJ1825 is
elongated along about the East--West direction. Following Plionis \&
Basilakos (\citeyear{plionis2002}, based on \citealt{carter1980}), we
computed the ellipticity ($\epsilon$) and the position angle of the
major axis ($\theta$). Using the 49 red galaxies at $R< 0.4$ \hh, we
obtained $\epsilon=0.23_{-0.11}^{+0.07}$ and $\theta=80_{-18}^{+15}$
deg.  This ENE-WSW direction agrees with the direction joining the two
BCGs and the elongation of the X-ray isophotes in the central region
(see Fig.~\ref{figottico} and \citealt{clavico2019}).  This concordance
is often found in clusters with evidence of past merger (e.g.,
\citealt{barrena2014}). Since the two BCGs have the same
line--of--sight velocity, their relative motion can only take place in the
plane of the sky.

The fact that the two BCGs themselves are strongly elongated may also
suggest that the axis connecting the two BCGs lies in the plane of the
sky. Moreover, they are both aligned with the cluster. The alignment
of dominant galaxies with the parent cluster has been reported by
numerous authors (see \citealt{joachimi2015} for a review), but in the
literature there is not a particular emphasis on clusters showing two
dominant galaxies. Indeed, the second brightest galaxy is found to be
very weakly aligned with the first one (e.g., \citealt{trevese1992};
\citealt{niederste2010}).  Two close galaxies are expected to become
aligned if they are influenced by the same gravitational forces, or if
the galactic cannibalism scenario has a preferred direction. In any
case, these mechanisms can be related also with the large scale
structure since dominant galaxies are found to be aligned with the
neighboring clusters on scales of several tens of \h (e.g.,
\citealt{west1994}) and the alignment of dominant galaxies is found to
be robust against major cluster mergers
(\citealt{wittman2019}). Therefore, the elongation of the two BCGs and
their alignment with the cluster structure are not useful to fix the
time when the two BCGs, and the likely related subclusters, joined
into forming the present cluster.

Clavico et al. (\citeyear{clavico2019}) claims that the ICM in the central part of
RXCJ1825 is going to relax. The inspection of the TNG image of the two
dominant galaxies can add some support to this point, in fact suggests
that there is an extra light in the region between the two BCGs (see
Fig.~\ref{figBCGs}). This intracluster light can be taken as an
evidence that the two galaxies have already interacted, that is the
real/deprojected velocity is small and that we are catching RXCJ1825
in a very advanced phase of formation.

Another trace of cluster assembly in RXCJ1825 is the detection of a
substructure at $\sim 0.4$ \h North--East of the center
(RXCJ1825NE). It is a poorly dense substructure with no peculiarity in
the velocity.  It is also detected in the analysis of red galaxies so
it might be a small merging group but also the remnant of a more
important past merger. Very interestingly, this substructure at
North--East and the SG galaxy discussed below trace the same direction
through the cluster center, thus suggesting that NE-SW may be another
direction of cluster accretion.

We also report the presence of an overdensity of high velocity
galaxies in the region between RXCJ1825 and CIZAJ1824
(MiddlePeak). This overdensity is mostly related to blue galaxies
(cfr. Figs.~\ref{figds} and \ref{figred} -- lower panel), thus
suggesting the presence of a few galaxies or a loose group just
infalling from the field onto the Lyra system and there projected.

\begin{figure}
\centering
\resizebox{\hsize}{!}{\includegraphics{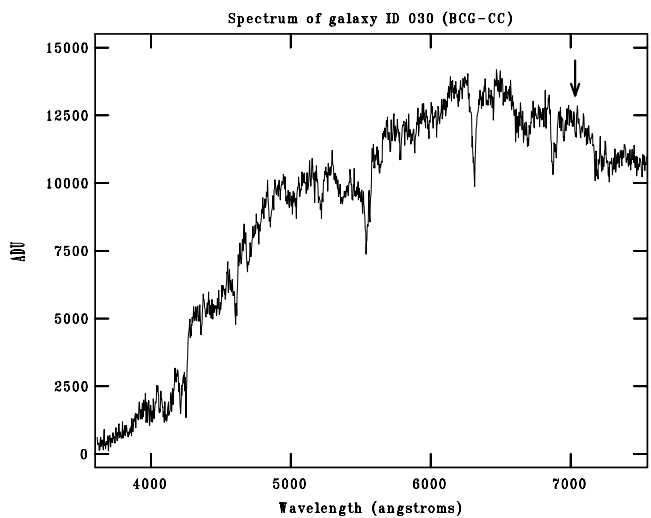}}
\caption
    {TNG spectrum of BCG-CC where no emission line is
      present, in particular H$\alpha$ expected at $\lambda
      \sim 7020$ angstrom at the cluster redshift. }
\label{figspec}
\end{figure}

As for CIZAJ1824, the BCG-CC is very luminous, with a magnitude value
intermediate between those of BCG-E and BCG-W. It is elongated in the
North--South direction, slightly NNE-SSW, in the same direction as the
X-ray isophotes (see Fig.~\ref{figottico}). Clavico et
al. (\citeyear{clavico2019}) report an entropy value of $16.1\pm0.3$
keV~cm$^{2}$ in the central region and one would expect that a BCG-CC
surrounded by this low entropy cool core were characterized by
H$\alpha$ emission (\citealt{cavagnolo2008}).  Instead, the spectrum
of BCG-CC does now show trace of emission lines, in particular no
H$\alpha$ line is detected (see Fig.~\ref{figspec}). Thus CIZAJ1824 is
one of the few clusters with low entropy and no H$\alpha$ emission, a
rare exception to the much larger trend (see Fig.~1 of
\citealt{cavagnolo2008}).

Our estimates of the velocity dispersion of RXCJ1825 and CIZAJ1824
galaxy populations are $\sigma_{\rm v}=995_{-125}^{+131}$ \ks and
$\sigma_{\rm v}=700\pm50$ \kss, respectively.  For each clusters,
$\sigma_{\rm v}$ can be compared with the estimates of the average
X--ray temperature, $kT=4.86\pm0.05$ keV and $kT=2.14\pm0.05$ keV
(\citealt{clavico2019}), computing the value of $\beta_{\rm
  spec}=\sigma_{\rm v}^2/(kT/\mu m_{\rm p})$, with $\mu=0.58$ the mean
molecular weight and $m_{\rm p}$ the proton mass.  We obtained
$\beta_{\rm spec}=1.2\pm0.3$ and $\beta_{\rm spec}=1.4\pm0.2$ for
RXCJ1825 and CIZAJ1824 respectively, that is a two-sigma agreement
with $\beta_{\rm spec}=1$, the equipartition of energy per unit mass
in galaxies and ICM (see also Figs.~\ref{figprof} and \ref{figprofCC},
bottom panels).

RXCJ1825 is found to be a massive cluster, $M_{200}=1.1\pm0.4$ \mqui
and CIZAJ1824 a low mass cluster, $M_{200}=4\pm1$ \mquaa. Our
$M_{200}$ mass estimates for RXCJ1825 and CIZAJ1824 are in agreement
with those based on X-ray data within uncertainties ($M_{200}=7\pm2$
\mqua and $M_{200}=4.2\pm1.5$ \mquaa, \citealt{clavico2019}).

\subsection{SG and RG in the SW region}
\label{discuSW}

The analysis of Clavico et al. (\citeyear{clavico2019}) shows a clear excess of diffuse
X-ray emission South--West of RXCJ1825, in the same direction of
SG. They suggest that this is related to a recent merger where a
group, now traced by its central galaxy SG, has interacted with
RXCJ1825 and is in an advanced state of disruption. This picture is
also well in agreement with the presence of an elongation of the radio
halo just in the same direction (\citealt{botteon2019}).
We find that the South--West region of the velocity field is   
very peculiar for its high velocity and the SG galaxy, there embedded,
is the fifth brightest galaxy of the Lyra field.

Notice that, although the South--West and CIZAJ1824 regions are
characterized by similar line--of--sight velocities, the geometry and
kinematics is likely to be completely different.  Assuming that SG is
outgoing from RXCJ1825, SG is on the back of RXCJ1825 running out from
RXCJ1825, while CIZAJ1824 is in the front and incoming.  Looking at
Fig.~\ref{figprof} one might doubt that SG is a galaxy really bound to
RXCJ1825. To better understand the fate of SG, we recomputed the
caustics for a toy-model cluster, that is that the ``future'' system
formed of RXCJ1825 and CIZAJ1824, where we consider the mass of the
whole system $M_{\rm sys}$.  SG lies well inside the caustics, that is
SG is likely to be bound to the Lyra system.

As for the search of possible companion galaxies of SG, our redshift
data poorly sample the South--West region and we success in detecting
a galaxy overdensity close to SG only when applying the 2D-DEDICA
analysis to photometric members (see Fig.~\ref{figk2cmri}).  We need
more redshift data around SG to be more conclusive.  However, the fact
that we fail in detecting a round, dense group of galaxies around SG
might be not so unexpected due to the disruption of the group itself.
In fact, it is well known that an interaction between two systems can
enhances the internal energy of the individual systems, which may
react losing part of their particles (\citealt{binney1987}, see
chapter 7.2).  In fact, from the observational side, there is a
plethora of features claimed to be the remnants of past mergers, such
as a group partially destroyed ejecting its central galaxy
(\citealt{colless1996}), a plume of outflying galaxies
(\citealt{flores2000}), a tidal debris stripped from the main cluster
(\citealt{owers2011}). Very likely, the mean velocity of the lost
companion galaxies should be about that of the barycenter, in our case
that of SG. These lost companions might be the cause of the high
velocity excess in the SW region.

Considering galaxies with velocities higher than that of SG, we detect
a stretched structure going from RXCJ1825 to SG (see
Fig.~\ref{figxyhv}). However, a very detailed analysis of X-ray data
suggests that two of these are just at the beginning of their
interaction with the ICM of RXCJ1825 and thus likely not part of the
same group of SG (see \citealt{clavico2019} for details).

Finally, we discuss RG, the tailed radio-galaxy lying South of
CIZAJ1824 and discovered by Botteon et
al. (\citeyear{botteon2019}). RG is characterized by a velocity
similar to that of BCG-CC and results well bound to the Lyra
complex. Due to the presence of the tails we expect that RG is just
interacting with the ICM, although it is not clear if the ICM of
CIZAJ1824 or RXCJ1825.  Herebelow we discuss two possible alternative
scenarios. The direction suggested by the tails points slightly at
West of RXCJ1825 (see \citealt{botteon2019}), thus RG might be related
to this cluster. In this case it is in front of RXCJ1825 running
toward it.  Alternatively, the pointing of its direction is not so
important and RG is moving toward CIZAJ1824 with a partially
tangential orbit. Indeed, the RG velocity is so close to the velocity
of BCG-CC, differing for only 230 \ks in the rest frame, that the
motion of RG is likely taking place mostly in the plane of sky.  In
this case the real distance of RG from the CIZAJ1824 center is similar
to the projected distance, $\sim 0.4$ \hh, and thus RG is embedded
inside the $R_{500}$ region of CIZAJ1824.  With present data we cannot
discriminate between the two scenarios and we suggest to think at RG
as a galaxy which is infalling toward the whole Lyra complex.

\section{Summary and conclusions}
\label{summa}

We present the first dynamical analysis of the Lyra
complex, formed of two clusters of galaxies RXCJ1825 and CIZAJ1824, as
based on the kinematics of member galaxies. New spectroscopic data for
285 galaxies were acquired at TNG and PanSTARRS magnitudes r, g, i
were used.  We selected 198 cluster members we used for most our
analyses. Herebelow we list our main results and conclusions:

   \begin{enumerate}
   \item
  Our analysis of the galaxy distribution well detects RXCJ1825 and
  CIZAJ1824 as individual units, at the distance of $\sim 16$ \arcm
  and points out that RXCJ1825 is more populated and more dense of
  CIZAJ1824 suggesting a mass ratio in the range from 2:1 to 4:1.
\item
     The redshifts of RXCJ1825 and CIZAJ1824 are $z=0.0645$ and
     $z=0.0708$.  We report the first estimates of velocity
     dispersion, $\sigma_{\rm v}=995_{-125}^{+131}$ \ks and
     $\sigma_{\rm v}=700\pm50$ \kss, for RXCJ1825 and CIZAJ1824,
     respectively. Following the recipe of \citet{munari2013}, our
     estimates of dynamical mass are  $M_{\rm 200}=1.1\pm
     0.4$ \mqui and $M_{\rm 200}=4\pm 0.1$ \mquaa, with a $\siml$ 3:1 mass
     ratio, in agreement with the point above.
   \item When assuming that cosmological distance is given by the
     redshift of the whole Lyra system, $z=0.0674$, the
     line--of--sight velocity difference and the projected distance
     between RXCJ1825 and CIZAJ1824 are $\sim 1750$ \ks and $D\sim
     1.3$ \hh. A dynamical analysis indicates that clusters are likely
     to be gravitationally bound and that CIZAJ1824 lies in front of
     RXCJ1825 and running toward it.  The specific case with the
     projection angle $\alpha=50$\degree\ leads to collision
     parameters quite reasonable, in particular the $R_{500}$-regions
     of the two clusters are just in touch but not compenetrated, in
     agreement with that no relevant interaction is detected in the
     X-ray data (\citealt{clavico2019}).  The future cluster is expected
     to be very massive, with a mass value in the 1.5-3 \mqui range.
     \item
       RXCJ1825 is found to be not relaxed and we have evidence that
       the likely merger related to the two dominant galaxies, a very
       past merger according to Clavico et
       al. (\citeyear{clavico2019}), lies in the plane of the sky
       along the East-West, slightly ENE-WSW, direction. The merger is
       also supported by the detection of a radio halo
       (\citealt{botteon2019}).
    \item
      The South--West region of the velocity field, where the
      high-velocity and luminous galaxy SG in embedded, is very
      peculiar for its high velocity, comparable to that of
      CIZAJ1824. We show that this is not expected in a model where
      there are only the two clusters. Rather, it suggests the
      presence of an additional population of high velocity
      galaxies. This supports the suggestion of Clavico et
      al. (\citeyear{clavico2019}) that SG was the central galaxy of a
      group just disrupted by its interaction with RXCJ1825, an
      interaction causing the elongation of X-ray isophotes toward the
      SW direction, the same direction of the extension of the radio
      halo (\citealt{botteon2019}).
   \end{enumerate}

      Our results shows that the Lyra region hosts a very complex,
      just assembling galaxy structure, in agreement with the picture
      delineated from recent radio and X-ray studies
      (\citealt{botteon2019}; \citealt{clavico2019}). From the optical
      side, new improvements can be obtained collecting redshifts for
      a larger field in such way to study RXCJ1825 at least out to its
      $R_{200}$ radius, adding many more redshifts in the South--West
      region, and acquiring spectra at higher resolution, thus using
      detailed spectral features to study the evolution of galaxies in
      this interesting environment.

\begin{acknowledgements}

We thank the referee for useful and constructive comments.  We thank
Andrea Botteon and Rossella Cassano for useful discussions and the
access to their LOFAR radio contours previous than the publication. We
thank Barbara Sartoris for an useful cosmological input.\\

  M.G. acknowledges financial support from the grant MIUR PRIN 2015
  ``Cosmology and Fundamental Physics: illuminating the Dark Universe
  with Euclid'' and from the University of Trieste through the program
 ``Finanziamento di Ateneo per progetti di ricerca scientifica - FRA
  2018''.\\

  This publication is based on observations made on the island of La
Palma with the Italian Telescopio Nazionale Galileo (TNG), which is
operated by the Fundaci\'on Galileo Galilei -- INAF (Istituto
Nazionale di Astrofisica) and is located in the Spanish Observatorio
of the Roque de Los Muchachos of the Instituto de Astrof\'isica de
Canarias.\\

This research has made use of the galaxy catalog of The Pan-STARRS
Survey (DR1) and its public science archive. They have been made
possible through contributions by the Institute for Astronomy, the
University of Hawaii, the Pan-STARRS Project Office, the Max-Planck
Society and its participating institutes, the Max Planck Institute for
Astronomy, Heidelberg and the Max Planck Institute for
Extraterrestrial Physics, Garching, The Johns Hopkins University,
Durham University, the University of Edinburgh, the Queen's University
Belfast, the Harvard-Smithsonian Center for Astrophysics, the Las
Cumbres Observatory Global Telescope Network Incorporated, the
National Central University of Taiwan, the Space Telescope Science
Institute, the National Aeronautics and Space Administration under
Grant No. NNX08AR22G issued through the Planetary Science Division of
the NASA Science Mission Directorate, the National Science Foundation
Grant No. AST-1238877, the University of Maryland, Eotvos Lorand
University (ELTE), the Los Alamos National Laboratory, and the Gordon
and Betty Moore Foundation.

\end{acknowledgements}

\bibliographystyle{aa}
\bibliography{biblio}

\end{document}